\documentclass[a4paper,12pt]{article}

\usepackage{graphicx}%
\usepackage{amsmath}
\usepackage{bm}
\usepackage{subcaption}
\usepackage{overcite}
\usepackage{mathrsfs}
\usepackage{lscape}
\usepackage{rotating}
\usepackage[version=3]{mhchem}
\usepackage{longtable}
\usepackage{setspace}
\usepackage{braket}

\title{Subsystem real-time Time Dependent Density Functional Theory}
\author{Alisa Krishtal$^1$, Davide Ceresoli$^{1,2}$ and Michele Pavanello$^1$}

\begin{document}

\section*{Copyright}

Copyright 2015 American Institute of Physics. This is a preprint version of an article submitted to The Journal of Chemical Physics. This article may be downloaded for
personal use only. Any other use requires prior permission of the author and the American Institute of Physics. \\
http://jcp.aip.org/

\newpage

\maketitle

\centerline{$^1$Department of Chemistry, Rutgers University, Newark, USA.}

\centerline{$^2$CNR-ISTM, Institute of Molecular Sciences and Technologies, Milan, Italy}

\begin{abstract}
We present the extension of Frozen Density Embedding (FDE) theory to real-time Time Dependent Density Functional Theory (rt-TDDFT). FDE a is DFT-in-DFT embedding method that 
allows 
to partition a larger Kohn-Sham system into a set of smaller, coupled Kohn-Sham systems. Additional to the computational advantage, 
FDE provides physical insight into the properties of embedded systems and the coupling interactions between them. The extension to rt-TDDFT is 
done straightforwardly by evolving the Kohn-Sham subsystems in time simultaneously, while updating the embedding potential between the systems at 
every time step. Two main applications are presented: the explicit excitation energy transfer in real time between subsystems is demonstrated for the 
case of the Na$_4$ cluster and the effect of the embedding on optical spectra of coupled chromophores. In particular, the importance of including the full dynamic response in 
the embedding potential is demonstrated.
\end{abstract}


\section{Introduction}

It is the nature of a chemist to view molecules as a collection of atoms and condensed states as a collection of molecules, and to interpret the properties of the larger 
aggregates in terms of their smaller building stones and the interactions between them. In the  field of computational chemistry, this tendency was first  expressed in a 
development of 
a range of so called Atoms-In-Molecule methods (AIM),\cite{bade1991,hirs1977,mull1955,beck1988,fons2004} which are meant to be used as a post-processing tool in order to reveal 
the 
subtle relations between structure and properties, hidden in the single quantity obtained from a supermolecular calculation. An array of AIM methods based on space 
partitioning,\cite{bade1991,beck1988,fons2004} density partitioning\cite{hirs1977} and wave function partitioning\cite{mull1955} have been developed over decades and successfully 
applied to a range of properties as charges, energies, response properties and reactivity analysis. Embedding\cite{weso1993,jaco2013,yang2002,wu2003,good2010,huang2011,cohe2007a} 
 is a more recently developed field which can be seen as taking AIM a step further and instead of applying its concepts in the post processing, merge it with the electronic 
structure theory. In this process, the goal to seek deeper chemical understanding has been supplemented by the aspiration 
to increase computational efficiency. Embedding has also opened the door to combining different electronic structure methods into a single calculation, building a bridge between 
the worlds of wave function theory and density functional theory (DFT).\cite{huang2011,good2010,csab2013,csab2014,weso2008,klun2001}

Subsystem DFT, also referred to as Frozen Density Embedding (FDE) theory (see Section \ref{sec:Subsystem_DFT} for discussion), is closely related to the Hirshfeld AIM 
method,\cite{hirs1977} as both view the total density at each point of space as a sum of subsystem densities
\begin{equation}
\label{eq:rho}
\rho_{tot}(\mathbf{r})=\sum_I^{N_S}\rho_{I}(\mathbf{r}).
\end{equation}
Where the Hirshfeld method handles each atom as a separate entity, FDE partitions the subsystem into molecular fragments.
Though Eq.\ (\ref{eq:rho}) does not seem to imply it at first sight, the partitioning also includes the division of 
space. The densities of the subsystems can overlap and, while they are in principle allowed to be delocalized over all space, in practice the 
initial guess is based on a chemically sensible representation which usually leads to a collection of weakly overlapping densities, centered around the nuclei of the 
subsystems.\cite{bern2008,humb2013} The``cutting'' of covalent bonds is avoided, since the description of the interaction between the subsystems relies on the use of 
approximate nonadditive kinetic energy functionals (NAKE), which, at this point, still 
struggle with strongly overlapping densities.\cite{silv2012} This through-space partitioning is the key to the simplification of the computational problem in FDE, which allows to 
take advantage of 
computational 
techniques such as nearsightedness and the use of localized basis sets. In this sense  FDE succeeds in combining 
the two goals of deeper chemical insight and computational efficiency.\cite{jaco2008b,weso1996b,weso1997c,geno2014,gonz2009,laha2007}

Most embedding methods are developed to be used for ground state calculations, though recently several have been extended to the realm of time dependent 
methods.\cite{casi2004,neug2007,mosq2013,huang2014,gome2008,gome2012} The first extensions have been made for the linear response formalism of time-dependent DFT 
(LR-TDDFT),\cite{casi2004,neug2007} a theory developed for the calculation of excited states that has increasingly gained in popularity in the last two 
decades.\cite{gisb1998,onid2002,marq2004} Real-time time-dependent density functional theory (rt-TDDFT), on the other hand, has only relatively recently came into the spot 
light,\cite{yaba1996,qian2006,cast2004,meng2010} mostly due to its role as a stepping stone into the world of nonadiabatic dynamics. In contrast to LR-TDDFT, rt-TDDFT does not 
rely on the 
linear response formalism and is intrinsically able to describe the full response of the system to an applied perturbation.\cite{marq2004} With the introduction of the Kohn-Sham 
DFT (KS-DFT) 
formalism and 
the adiabatic approximation (see Section \ref{subsystemrtTDDFT} for discussion) it reduces to a non-linear integration in time of the time dependent Kohn-Sham (TDKS) equations. 
While 
this can be computationally demanding, as it requires very small time steps \big[$\sim$ 1-2 attoseconds (as)\big] and long simulation times \big[$\sim$ 10-200 femtoseconds 
(fs) depending on the
application\big], it has the advantage of scaling as $\mathcal{O}(N^3)$. While this implies that linear scaling techniques can be applied as in the case of ground state KS, one 
must be 
cautious with introducing numerical errors as they can accumulate during the long simulation process.

The integration of rt-TDDFT in embedding methods has only started recently and, 
until this work, was limited to the exact embedding methods\cite{mosq2013,huang2014} The spirit of these methods is very different from FDE, as they require a precalculated
total density, aiming at combining different 
computational methods. Here we present a new formulation of  subsystem rt-TDDFT and show (see Section \ref{eet} for discussion) that combining 
rt-TDDFT with a subsystem approach can offer a view into the process of excitation energy transfer, which cannot be obtained straightforwardly from standard rt-TDDFT 
calculation.\cite{hofm2010} Additionally, subsystem rt-TDDFT readily provides information about the role of coupling between the subsystems in the reproduction of optical 
absorption 
spectra, a question that has been raised before in LR-TDDFT applications.\cite{neug2007}


\section{Method}
\subsection{Subsystem DFT}
\label{sec:Subsystem_DFT}

In this section we will review the fundamentals of subsystem DFT before extending the formalism to rt-TDDFT. In subsystem DFT the partitioning of 
the total system into subsystems is made at the density level [Eq.\ (\ref{eq:rho})]. In DFT, the density is the central quantity and is used to formulate the variational principle 
for 
the ground state energy of the system
\begin{equation}
 E_0=\min_{\rho\rightarrow N}E[\rho]=\min_{\rho \rightarrow N}\left( F[\rho] + \int v_{ext}(\mathbf{r})\rho(\mathbf{r}) \right).
\end{equation}
where, using Levy's constrained search, $\rho$ is restricted to be $N$-representable, $v_{ext}(\mathbf{r})$ is the external potential due to the electron-nuclei interaction and 
$F[\rho]$ 
is the universal functional. $F[\rho]$ is defined as
\begin{equation}
\label{eq:F}
 F[\rho] = T[\rho] + W[\rho] = T_s[\rho] + J[\rho] + E_{xc}[\rho]
\end{equation}
where $T[\rho]$ is the kinetic energy and $W[\rho]$ is the electron-electron interaction. $F[\rho]$ can also be rewritten using 
the noninteracting kinetic 
energy $T_s[\rho]$, the Coulomb energy $J[\rho]$ and the exchange-correlation energy $E_{xc}[\rho]$ , defined as the difference between $T[\rho]$ and $T_s[\rho]$ 
and $W[\rho]$ and 
$J[\rho]$.
In the case of KS-DFT, the density $\rho$ is mapped to a set of $N$ non-interacting electrons represented by a single Slater determinant consisting out 
of the KS orbitals $\phi_i(\mathbf{r})$ and is given, for closed-shell systems, by
\begin{equation}
 \rho(\mathbf{r})=\sum_i 2 |\phi_i(\mathbf{r})|^2
\end{equation} 
This allows to rewrite the energy expression as 
\begin{equation}
 E_0=\min_{\phi\rightarrow \rho \rightarrow N}(E[\rho])=\min_{\phi}\left(T_s[\{\phi_i\}]+ \int v_{eff}(\mathbf{r})\rho(\mathbf{r})\right).
\end{equation}
where the noninteracting kinetic energy is now expressed exactly using the KS orbitals
\begin{equation}
 \label{eq:Ts}
 T_s[\{\phi_i\}] = \sum_{i=1}^{N} \Braket{\phi_i |-\frac{1}{2}\nabla^2 | \phi_i},
\end{equation}
and all electron-electron interaction terms are stacked away in an \textit{effective} potential 
\begin{equation}
 v_{eff}(\mathbf{r}) = v_{ext}(\mathbf{r}) + \frac{\delta J[\rho]}{\delta \rho(\mathbf{r})} + \frac{\delta E_{xc}[\rho]}{\delta \rho(\mathbf{r})}
\end{equation}
Minimizing $E[\rho]$ with respect to the KS orbitals is thus achieved by
\begin{equation}
 \frac{\delta}{\delta \phi^*_k}\left[ E[\rho] - 
\sum_{ij}^N\epsilon_{ij}\left(\int\phi_i^*(\mathbf{r})\phi_j(\mathbf{r})d\mathbf{r}-\delta_{ij}\right)\right]=0, ~ \forall k
\end{equation}
which leads to the one-electron KS equations
\begin{equation}
 \left[-\frac{1}{2}\nabla^2+v_{eff}(\mathbf{r})\right]\phi_i(\mathbf{r})=\epsilon_i\phi_i(\mathbf{r})
\end{equation}

When the total density is decomposed into a sum of subsystem densities $\{\rho_I(\mathbf{r})\}$ [Eq.\ (\ref{eq:rho})], the total energy functional $E[\rho]$ can be rewritten as
\begin{equation}
\label{eq:e}
 E[\rho]=E[\{\rho_I\}]=\sum_I\int 
v_{ext}(\mathbf{r})\rho_I(\mathbf{r})d\mathbf{r}+\sum_I\sum_J\int\int\frac{\rho_I(\mathbf{r})\rho_J(\mathbf{r}')}{|\mathbf{r}-\mathbf{r}'|}d\mathbf{r}
d\mathbf { r } '+T_s \left[ \sum_I\rho_I \right]+E_{xc}\left[\sum_I\rho_I\right],
\end{equation}
where the external potential term is additive, the Coulomb repulsion term is pairwise additive and the kinetic energy and exchange correlation energy 
are nonadditive. The nonadditive density functionals can be rewritten as
\begin{align}
\label{eq:K}
 K\left[\sum_I\rho_I\right]&= \sum_IK\left[\rho_I\right] + \left(K\left[\rho\right]-\sum_IK\left[\rho_I\right]\right) = \sum_IK\left[\rho_I\right] + K^{nadd}\left[\{\rho_I\}\right]
\end{align}
If the division into subsystem densities is done in such a way that each one 
is \textit{noninteracting v}-representable, then they can be viewed as separate KS subsystems, each mapped to an effective KS potential and a set of subsystem KS 
orbitals $\{\phi^I_i(\mathbf{r})\}$. The additive part of the noninteracting kinetic energy functional in Eq.\ 
(\ref{eq:K}) can then be expressed exactly using the 
subsystem orbitals $T_s[\{\phi^I_i\}]= \sum_{i=1}^{N_I} \braket{\phi_i^I | -\frac{1}{2}\nabla^2 | \phi_i^I}$, while the nonadditive part is expressed using a 
kinetic energy functional $\tilde{T}_s[\{\rho_I\}]$. In the spirit of KS, all other terms in Eq.\ (\ref{eq:e}) are gathered together in the subsystem effective potential 
$v^I_{eff}(\mathbf{r})$ after taking a functional derivative to $\rho_I$
\begin{equation}
\label{eq:veffI}
 v_{eff}^I(\mathbf{r})=v_{ext}(\mathbf{r})+\frac{\delta J[\rho]}{\delta\rho(\mathbf{r})} + \frac{\delta E_{xc}[\rho]}{\delta \rho(\mathbf{r})} + 
\frac{\delta\tilde{T}_s[\rho]}{\delta\rho(\mathbf{r})}-\frac{\delta\tilde{T}_s[\rho_I]}{\delta\rho_I(\mathbf{r})}
\end{equation}
Minimizing the energy of the total system is then replaced by a set of coupled minimizations
\begin{equation}
\label{eq:FDEmin}
 E_0=\min_{\{\phi^1_i\}}\min_{\{\phi^2_i\}}...\min_{\{\phi^{N_I}_i\}}(E[\{\rho_I\}]).
\end{equation}
Each of the minimizations is equivalent to
\begin{equation}
\frac{\delta}{\delta \phi^{I*}_k}\left( E[\{\rho_I\}] - 
\sum_{ij}\epsilon_{ij}^I\left[\int\phi_i^{*I}(\mathbf{r})\phi^{I}_j(\mathbf{r})d\mathbf{r}-\delta_{ij}\right]\right)=0
\end{equation}
which leads to a set of coupled subsystem KS equations
\begin{equation}
\label{eq:coupledKS}
 [-\frac{1}{2}\nabla^2+v_{eff}^I(\mathbf{r})]\phi_i^I(\mathbf{r})=\epsilon_i^I\phi_i^I(\mathbf{r})
\end{equation}

Derivation of the subsystem KS equations depends on the following constraints for each subsystem $I$:
\begin{enumerate}
 \item \label{constraint1}The density is \textit{noninteracting pure-state v-representable} ($v_s$-representable).
 \item \label{constraint2} $\delta \rho_{J}=0, J \ne I$, to insure that $\frac{\delta \rho(\mathbf{r})}{\delta \rho_I(\mathbf{r}')}=\delta(\mathbf{r}-\mathbf{r'})$ 
and $\frac{\delta \rho_I(\mathbf{r})}{\delta \rho_J(\mathbf{r}')}=\delta_{IJ}\delta(\mathbf{r}-\mathbf{r'})$.
\end{enumerate}
Due to constraint \ref{constraint2}, Eqs.\ (\ref{eq:coupledKS}) are solved iteratively through ``freeze-and-thaw'' cycles,\cite{weso1996b} where the energy is 
minimized with respect to subsystem $I$ while keeping the densities of all other subsystems frozen. At self consistency, the energy is minimized 
with respect to the variation of all subsystem densities and therefore also the total density. 

The subsystem effective potential in Eq.\ (\ref{eq:veffI}) can be rewritten as the KS potential of the total system plus the nonadditive kinetic energy potential
\begin{equation}
 v_{eff}^I(\mathbf{r})=v_{eff}^{tot}(\mathbf{r})+
\frac{\delta\tilde{T}_s[\rho]}{\delta\rho(\mathbf{r})}-\frac{\delta\tilde{T}_s[\rho_I]}{\delta\rho_I(\mathbf{r})}
\end{equation}
where we use the notation $\tilde{T}_s$ to emphasize that an approximate density functional is used. In case of the exact kinetic energy functional $T_s[\rho]$
the subsystem effective KS potential would reduce to the exact KS potential associated with the embedded density $\rho_I(\mathbf{r})$ plus a constant
\begin{equation}
  v_{eff}^I(\mathbf{r})= v_{eff}^{KS,I}(\mathbf{r}) + \mathrm{const}
\end{equation}
where we used the relationship\cite{ayer2004}
\begin{equation}
 \frac{\delta T_s[\rho]}{\delta\rho(\mathbf{r})} = \mathrm{const} - v_{eff}^{KS}(\mathbf{r})
\end{equation}
In other words, in the limit of exact NAKE, FDE reproduces the exact density and KS orbitals of the embedded fragment. 
However, in practice, approximate
NAKE functionals are used for the nonadditive kinetic energy and the subsystem KS effective potential differs from the
the exact KS potential by\cite{grit2013}
\begin{equation}
v_{eff}^{I}(\mathbf{r})-v_{eff}^{I,KS}(\mathbf{r})=\frac{\delta T_s[\rho]}{\delta\rho(\mathbf{r})}-\frac{\delta\tilde{T}_s[\rho]}{\delta\rho(\mathbf{r})}+
\frac{\delta T_s[\rho_I]}{\delta\rho_I(\mathbf{r})}-\frac{\delta\tilde{T}_s[\rho_I]}{\delta\rho_I(\mathbf{r})}
\end{equation}
The performance of the approximate NAKE is therefore essential for the reproduction of the densities of the embedded fragment and, as a result, also the total supermolecular 
density. 

During the freeze-and-thaw (FAT) cycles, the density of subsystem $I$ is being repeatedly optimized, each time in the 
presence of a different frozen density constraint. At each FAT cycle, the self consistency corresponding to a different KS potential 
$v_{eff}^I(\mathbf{r})$ is found. This process  can be considerably sped up by updating the KS potential after each 
SCF cycle.\cite{jaco2008b} Since the self-consistency point is well defined, both approaches lead to the same result.

Subsystem DFT has been shown to 
perform well for 
systems where the density can be partitioned between non-covalently bonded systems, i.e., systems interacting through electrostatics, van der Waals 
forces and hydrogen bonds.\cite{weso2001,weso2003,weso2004,kiew2008} Covalently bonded fragments, as well as systems with partial charge transfer character, remain a challenge 
for subsystem 
DFT until more accurate NAKE are developed.\cite{fux2010} When applied to systems with clearly separated fragments,  subsystem DFT offers the advantage of 
providing us with a chemically 
sensible approximation of the density and properties of \textit{embedded} systems. 
The embedding potential, given below, represents the difference between the KS potential of the noninteracting fragment, $v_{noint}^I$ (i.e., evaluated with the density of the 
embedded fragment but not including the interaction with the environment), and the effective KS 
potential of the embedded subsystem $v_{eff}^I$. As a result, it contains the full information about the interaction and coupling between the 
subsystems.
\begin{equation}
\label{eq:vemb}
 v_{emb}^I(\mathbf{r}) = v_{eff}^I(\mathbf{r}) - v_{noint}^I(\mathbf{r}) = \sum_{J \ne I}v_{ext}^J(\mathbf{r})+ \sum_{J \ne 
I}\int\frac{\rho_J(\mathbf{r})}{|\mathbf{r}-\mathbf{r}'|}d\mathbf{r}' + \frac{\delta E^{nadd}_{xc}}{\delta \rho_I(\mathbf{r})} +  \frac{\delta 
\tilde{T}_s^{nadd}}{\delta \rho_I(\mathbf{r})}
\end{equation}

\subsection{Subsystem real time TDDFT}
\label{subsystemrtTDDFT}

The central point of the derivation of subsystem DFT for the ground state is the variational principle, i.e., minimization of the total energy of 
the system with respect to the density of each subsystem. In contrast to the ground state case, there is no such minimum principle for the time 
dependent problem and therefore no unique way to derive the time evolution of a system. A possible route is to use the stationary 
action principle which states that the variation of the quantum mechanical action $\mathcal{A}$, defined as:
\begin{equation}
\label{eq:A}
 \mathcal{A}[\Psi]=\int_{t_0}^{t_1}dt \Braket{\Psi(t) | i\frac{\partial}{\partial t} - \hat{H}(t) | \Psi(t)}
\end{equation}
should be zero. In Eq.\ (\ref{eq:A}), $\hat{H}$ is the full molecular Hamiltonian that includes the time-dependent applied potential and $\Psi(t)$ is the electronic wavefunction. 
By imposing the boundary conditions that $\delta \Psi(t_0)=\delta 
\Psi(t_1)=0$ one recovers the time-dependent Schr\"odinger equation
\begin{equation}
 i\frac{\partial}{\partial t}\Psi(t)=\hat{H}(t)\Psi(t)
\end{equation}

Since the introduction of the Runge-Gross theorem,\cite{rung1984} there has been ongoing discussion whether the stationary action principle can be carried over
straightforwardly to TDDFT. In TDDFT, the (fully interacting) quantum state $\Psi(t)$ is a functional of the time dependent density $\rho(\mathbf{r},t)$, making 
the action a density 
functional as well. 
\begin{align}
 \mathcal{A}[\rho]&=\int_{t_0}^{t_1}dt \Braket{\Psi[\rho](t) | i\frac{\partial}{\partial t} - \hat{H}(t) | \Psi[\rho](t)} \\
 &=\int_{t_0}^{t_1}dt  \Braket{\Psi[\rho](t) | i\frac{\partial}{\partial t} - \hat{T} - \hat{W} | \Psi[\rho](t)} - 
\int_{t_0}^{t_1}dt\int 
d\mathbf{r}\rho(\mathbf{r},t)v_{ext}(\mathbf{r},t)
\end{align}
As was shown by van Leeuwen,\cite{leeu1999,leeu2001,leeu2005} the density $\rho(\mathbf{r},t)$ which evolves in an interacting system under the influence of an external 
potential $v_{ext}(\mathbf{r},t)$ starting from an initial state 
$\Psi[\rho](t_0)$ can be reproduced in a KS noninteracting system evolving under 
the action of an effective potential $v_s(\mathbf{r},t)$ starting from an initial state $\Phi[\rho](t_0)$, where $\Phi[\rho]$ is the single Slater 
determinant representing the noninteracting system.
\begin{align}
\label{Aq}
 \mathcal{A}_s[\rho] &= \int_{t_0}^{t_1}dt \Braket{\Phi[\rho](t) | i\frac{\partial}{\partial t} - \hat{H}_s(t) | \Phi[\rho](t)} \notag \\
 &= \int_{t_0}^{t_1}dt \Braket{\Phi[\rho](t) | i\frac{\partial}{\partial t} - \hat{T}_s | \Phi[\rho](t)} - 
\int_{t_0}^{t_1}dt\int 
d\mathbf{r}\rho(\mathbf{r},t)v_s(\mathbf{r},t)
\end{align}
The difference $v_s(\mathbf{r},t) - v_H(\mathbf{r},t) - v_{ext}(\mathbf{r},t)$ defines the time dependent exchange-correlation potential 
$v_{xc}(\mathbf{r},t)$. The term $v_{ext}(\mathbf{r},t)$ contains here both the nucleus-electron interaction as well as any applied time-dependent field in the form of
\begin{equation}
 v_{appl}(\mathbf{r},t) = v_0(\mathbf{r})+v_{1}(\mathbf{r},t)\theta(t-t_0)
\end{equation}
where $\theta(t-t_0)$ represents a step function.

The variation of the TDDFT action is unfortunately not as simple a matter as in the wave function formulation of the theory. Since the action is a 
density functional, one needs to take the variation with respect to the density. The boundary conditions are, however, not transferable. One can 
still require $\frac{\delta \Phi[\rho]}{\delta \rho(\mathbf{r},t)}$ to be zero for $t_0$ but not for $t_1$, since a variation in $\rho$ at a time 
$t_0 < t < t_1$ can and will change the quantum state $\Phi[\rho]$ at time $t_1$. The (not longer that) stationary action principle can then be 
rewritten 
as
\begin{equation}
 \delta \mathcal{A}[\rho] - i \Braket{\Psi[\rho](t_1) | \delta \Psi[\rho](t_1)} =  \delta \mathcal{A}_s[\rho] - i \Braket{\Phi[\rho](t_1) | \delta 
\Phi[\rho](t_1)} = 0
\end{equation}
The problem is circumvented in the usual spirit of KS-DFT by shifting the unknown terms into the exchange-correlation potential, explicitly defined 
as
\begin{equation}
\label{eq:xct}
v_{xc}(\mathbf{r},t) = \frac{\delta A_{xc}[\rho]}{\delta\rho(\mathbf{r},t)} + i \frac{\Braket{\Psi[\rho](t_1) | \delta \Psi[\rho](t_1)}}{\delta\rho(\mathbf{r},t)}- i 
\frac{\Braket{ 
\Phi[\rho](t_1) | \delta \Phi[\rho](t_1)}}{\delta\rho(\mathbf{r},t)}
\end{equation}
In practical calculation, one usually resorts to the adiabatic approximation, which simplifies the time-dependent exchange-correlation 
potential to the ground state exchange correlation potential evaluated at the time dependent density. 
\begin{equation}
 v_{xc}^{adiabatic}(\mathbf{r},t)\equiv v_{xc}(\mathbf{r})[\rho(\mathbf{r},t)]=\left.\frac{\delta E_{xc}[\rho]}{\delta \rho}\right|_{\rho=\rho(\mathbf{r},t)}
\end{equation}
As a consequence, the extra terms in Eq.\ (\ref{eq:xct}) vanish, the density time dependence is reduced to instantaneous and causality is trivially 
fulfilled. The familiar one-electron time dependent KS equations (TDKS) take the form
\begin{equation}
 [-\frac{1}{2}\nabla^2+v_s(\mathbf{r},t)]\phi_i(\mathbf{r},t)=i\frac{\partial}{\partial t}\phi_i(\mathbf{r},t)
\end{equation}
where $\phi_i(\mathbf{r},t)$ are the noninteracting KS one-electron orbitals constituting the single Slater determinant $\Phi[\mathbf{\rho}](t)$ and yielding the time dependent 
density $\rho(\mathbf{r},t)$ 

Once a solution to the ground state coupled subsystem KS equations [Eq.\ (\ref{eq:coupledKS})] has been obtained, we can define an action 
principle for each of the subsystems $I$. Each subsystem is represented by a single Slater determinant $\Phi[\rho_I]$, which is a functional of 
the subsystem density $\rho_I(\mathbf{r},t)$. Namely,
\begin{align}
\label{eq:AI}
 \mathcal{A}^I_s[\rho_I] =& \int_{t_0}^{t_1}dt \Braket{\Phi[\rho_I](t) | i\frac{\partial}{\partial t} - \hat{T}_s | 
\Phi[\rho_I](t)} - \int_{t_0}^{t_1}dt\int d\mathbf{r}\rho_I(\mathbf{r},t)v_{s}^{I}(\mathbf{r},t),
\end{align}
where, in accordance with the adiabatic approximation, $v_{s}^I(\mathbf{r},t)$ is defined as:
\begin{equation}
 \label{eq:vsIrt}
 v_{s}^I(\mathbf{r},t)=v_{ext}(\mathbf{r},t)+\frac{\delta J[\rho]}{\delta\rho(\mathbf{r},t)} + \frac{\delta E_{xc}[\rho]}{\delta \rho(\mathbf{r},t)} 
+ 
\frac{\delta\tilde{T}_s[\rho]}{\delta\rho(\mathbf{r},t)}-\frac{\delta\tilde{T}_s[\rho_I]}{\delta\rho_I(\mathbf{r},t)}
\end{equation}
and is equal at $t=t_0$ to the $v_{eff}^I(\mathbf{r})$ in Eq.\ (\ref{eq:coupledKS}) at self consistency. We note that only $v_{ext}(\mathbf{r},t)$ contains an 
explicit time dependence while the other terms depend on time through the density.
In the limit case of exact NAKE, the sum of the subsystem actions [Eq.\ 
(\ref{eq:AI})] will reduce to the total KS action [Eq.\ (\ref{Aq})].
Taking the variation of each subsystem action allows us to obtain the subsystem TDKS equations
\begin{equation}
\label{eq:tdks}
 [-\frac{1}{2}\nabla^2+v_s^I(\mathbf{r},t)]\phi_i^I(\mathbf{r},t)=i\frac{\partial}{\partial t}\phi_i^I(\mathbf{r},t)
\end{equation}

In order to reproduce the total KS action, it is imperative to integrate the subsystem TDKS equations 
\textit{simultaneously}. The subsystem action, as defined in Eq.\ (\ref{eq:AI}), represents the \textit{coupled} action, since the subsystem 
potential $v_s^I(\mathbf{r},t)$ depends on the total time dependent density $\rho(\mathbf{r},t)=\sum_I\rho_I(\mathbf{r},t)$ of the system. If the applied time dependent potential 
to the system is sufficiently small, one can draw a parallel to 
 the subsystem formulation of linear response TDDFT, where the coupled response function of the subsystem $\chi^{c}_I=\frac{\delta 
\rho_I(\mathbf{r},t)}{\delta v_{ext}(\mathbf{r}',t')}$, is connected to the 
coupled response functions of all the other subsystems through a Dyson-type equation (in simplified notation)\cite{neug2007,pava2013b} 
\begin{equation}
 \label{eq:dyson}
 \chi^c_I = \chi^u_I+\sum_{J\ne I}^{N_s}\chi^u_IK_{IJ}\chi_J^c
\end{equation}
where $K_{IJ}=\frac{\delta v_{emb}^I(\mathbf{r},t)}{\delta \rho_J(\mathbf{r}',t')}$.
As in the case of subsystem formulation of linear response TDDFT, one can also define an equivalent to the uncoupled response of the subsystem $\chi^u_I$\cite{casi2004,pava2013b} 
\begin{equation}
 \chi_I^u=\chi_I^0+\chi_I^0K_{II}\chi_I^u
\end{equation}
where $\chi^0_I$ is the subsystem KS response function. This is achieved 
by defining the 
uncoupled subsystem action, where the density of the other 
subsystems are kept frozen and  Eq.\ (\ref{eq:tdks}) is only evolved in time for the subsystem in question.  
As we will show in Section \ref{sec:opt}, examining the 
difference in subsystem properties between the coupled calculations and the uncoupled calculation reveals first hand information on the 
excitation
coupling between the subsystems. It is also important to note that the embedding potential is never 
fully static also in the uncoupled calculations, as it depends on the total electron density which varies in time 
even when 
the other subsystems are kept frozen.

\section{Implementation details}
The subsystem rt-TDDFT method was implemented into the {\sc Quantum Espresso} package as an extension of the periodic subsystem DFT\cite{geno2014} using plane waves, the details 
of which can be found in Ref.\  \citenum{geno2014}. In this work we only report calculations involving molecular systems using the $\Gamma$ point sampling for the 
Brillouin 
zone. 
 All calculations have been performed using  
ultrasoft pseudopotentials\cite{vand1990,laas1991,laas1993}. Ultrasoft pseudopotentials can be seen as a 
special case of the projector 
augmented wave (PAW) method\cite{bloe1994} which reformulates the KS equations in terms of auxiliary smooth functions $\phi_i^{PS}$ to the true one-electron KS orbitals 
through a 
linear transformation operator $\hat{T}$. As a result, the orthogonality of the original KS orbitals is replaced by the orthogonality of the pseudo functions with respect to the 
overlap operator $\hat{S}=\hat{T}^{\dagger}\hat{T}$
\begin{equation}
\braket{\phi_i|\phi_j}= \braket{\phi^{PS}_i|\hat{S}|\phi^{PS}_j},
\end{equation}
and the expectation value of any local operator is given by
\begin{align}
 \braket{\hat{O}} &= \braket{\phi_i|\hat{O}|\phi_i} = \braket{\phi_i^{PS}|\hat{T}^{\dagger}\hat{O}\hat{T}|\phi_i^{PS}}.
\end{align}
The KS eigenvalue equation transforms then into a generalized eigenvalue equation
\begin{equation}
 \hat{H}_s^{PS}\ket{\phi_i^{PS}}=\epsilon_i\hat{S}\ket{\phi_i^{PS}}
\end{equation}
where the transformed Hamiltonian $\hat{H}_s^{PS}=\hat{T}^{\dagger}\hat{H}_s\hat{T}$ contains extra local and non-local pseudopotential projectors. In the 
time-dependent case, also the $i\frac{\partial}{\partial t}$ operator needs to be transformed
\begin{align}
 \hat{H}_s^{PS}\ket{\phi_i^{PS}}&=\hat{T}^{\dagger}i\frac{\partial}{\partial t}\hat{T}\ket{\phi_i^{PS}}
\end{align}
which leads to 
\begin{equation}
\label{ultratdks}
 (\hat{H}_s^{PS}+\hat{P})\ket{\phi_i^{PS}}=i\hat{S}\frac{\partial}{\partial t}\ket{\phi_i^{PS}}
\end{equation}
The additional operator $\hat{P}=-i\hat{T}^{\dagger}\frac{\partial \hat{T}}{\partial t}$ depends on the velocities of the nuclei and for TDDFT calculations with immobile ions is 
identical to zero. When integrating the rt-TDDFT method into Ehrenfest dynamics calculations with mobile atoms, explicit calculation of the $\hat{P}$ operator is 
necessary.\cite{qian2006,ojan2012} 

For the time propagation of Eq.\ (\ref{ultratdks}), we use the first-order Crank-Nicolson method,\cite{crank1996} where the KS orbitals are calculated at each time step from
\begin{equation}
\label{crank}
 \left[\hat{S}+\frac{i}{2}\hat{H}_s^{PS}\left(t+\frac{\Delta t}{2}\right)\Delta t\right]\phi_i^{PS}(t+\Delta t) =  \left[\hat{S}-\frac{i}{2}\hat{H}_s^{PS}\left(t+\frac{\Delta 
t}{2}\right)\Delta 
t\right]\phi_i^{PS}(t).
\end{equation}
Since we are using plane wave basis sets, the dimension of $\hat{H}_s(t)$ and $\hat{S}$ matrices is usually too large to allow a numerically stable inversion and Eq.\ 
(\ref{crank}) 
is 
solved using the Conjugated Gradient Square method. The advantage of the Crank-Nicolson method for the time propagation is the explicit  norm conservation of the wave 
function. Since we are using the first order version of the method, we forgo the predictor-corrector step and approximate $\hat{H}_s(t+\frac{\Delta t}{2} )$ by 
$\hat{H}_s(t)$. Thus, very small time steps must be used and 
in all calculations presented here we have used the $2$ attoseconds.


\section{Results and Discussion}
\label{sec:RandD}

\subsection{Optical spectra}

\label{sec:opt}

As a first application of the novel method, we will consider the calculation of an optical spectrum between two coupled chromophores. An optical spectrum can be 
obtained by applying a time dependent electric field to the system such that it samples all frequencies. In order to sample all frequencies equivalently, the electric field has 
ideally the form of a $\delta$-function, $E(t) \propto \delta(t-t_0)$. This can be done in two ways: one is adding an 
electric field in the form of a very narrow gaussian. The potential that is added 
to the periodic Hamiltonian of the system and multiplied in real space with a saw function, which drops from 1 to 0 in the part of the supercell furthest away from the 
nuclei. The second way is particularly suited for periodic systems: it prescribes to multiply the occupied KS states by a phase\cite{yaba1996}
\begin{equation}
\label{eq:pulse}
 \phi_i(\mathbf{r},t=0^{+})=e^{i\mathbf{E}\cdot\mathbf{r}}\phi_i(\mathbf{r},t=0^{-})
\end{equation}
where $\mathbf{E}$ is the field.
For molecular calculations, both methods result in identical time evolution of the dipole moment provided that the gaussian pulse integrates to the field strength $|E|$.
The dipole moment is calculated at each time step using
\begin{equation}
 \bm{\mu}(t)=\int \rho(\mathbf{r},t)\mathbf{r}d\mathbf{r}
\end{equation}
The oscillator strength is obtained by a Fourier transformation of the time-dependent dipole moment into the
frequency domain.
\begin{equation}
 S(\omega) = \frac{1}{3}\frac{\omega}{\pi}\sum_{km}Im[\alpha_{kk}(\omega)]
\end{equation}
\begin{equation}
 Im[\alpha_{kk}(\omega)]=-\frac{2}{E_k}\int \sin(\omega t)e^{-\gamma t^2}[\mu_{k}(t)-\mu_{k}(t_0)]dt
\end{equation}
where $k$ stands for one of the Cartesian directions $x,y$ or $z$ and $\gamma$ is a small damping factor.\cite{qian2006} We choose here a Gaussian damping function rather than 
the usual exponential as it produces cleaner spectra. As a result, the shape of the spectral bands is no longer Lorentzians. In a subsystem calculation, the field is applied to 
each 
subsystem and the total optical spectrum of the system is obtained by summing 
the oscillator strengths of the subsystems because
\begin{equation}
\rho(\mathbf{r},t)=\sum_I\rho_I(\mathbf{r},t) \Rightarrow \bm{\alpha}(\omega)=\sum_I\bm{\alpha}_I(\omega)  \Rightarrow  S(\omega)=\sum_I S_I(\omega)
\end{equation}

\begin{figure}[htp]
\centering
\caption[]{Geometry of the Benzene-Fulvene parallel stacked dimer}
 \includegraphics[scale=0.4]{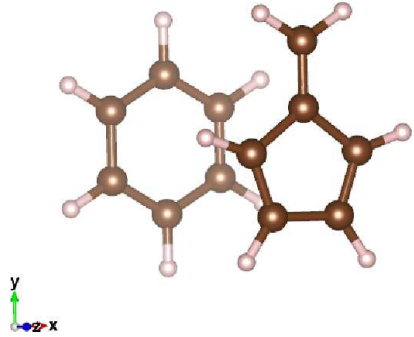}
  \label{fig:benzene-fulvene}
\end{figure}

We have calculated the optical spectra of a parallel-stacked benzene fulvene dimer (Fig.\ \ref{fig:benzene-fulvene}). The dimers were constructed by placing the relaxed monomers
along the Z-axis with an intermolecular distance of $4$, $5$, $6$, $7$ and $8$ \AA. The PBE functional\cite{PBEc} and the ultrasoft pseudopotentials from the GBRV 
library\cite{garri2014} where used 
throughout. For all calculations, 
a supercell of dimensions $31.0 \times 32.5 \times 37.8$ a.u.$^{3}$ was used with a kinetic energy cutoff of $55.0$ Ry and density cutoff of $600.0$ Ry. Each rt-TDDFT calculation 
was run for $16$ femtoseconds with a time step of $2$ attoseconds. The simulation time was tested for convergence by performing the Fourier transformation after each 
femtosecond until no differences in the resulting optical spectra were observed. The isolated monomers were calculated using the same supercell and cutoff. In the subsystem 
calculations, the LC94\cite{LC94} 
functional for the nonadditive kinetic energy was used throughout.

Table \ref{tab:eint} lists the interaction energy in the dimer obtained using KS-DFT and subsystem DFT for the range of separation distances and the number of electrons misplaced 
by FDE, $\braket{\Delta\rho}$ defined as\cite{jaco2008,kiew2008}
 \begin{equation}
 \label{eq:deltarho}
 \braket{\Delta\rho} = \frac{1}{2} \int \left| \Delta\rho(\mathbf{r}) \right| \, d\mathbf{r}
\end{equation}
where
\begin{equation}
 \Delta\rho(\mathbf{r}) = \rho_\text{FDE}(\mathbf{r}) - \rho_\text{KS}(\mathbf{r})
\end{equation}
\begin{table}[htp]
 \caption[Interaction energies]{The interaction energy in the benzene-fulvene dimer calculated using KS-DFT and subsystem DFT and the number of misplaced electrons [Eq.\ 
(\ref{eq:deltarho})] for the different intermolecular distances}
 \label{tab:eint}
 \centering
 \begin{tabular}{lccc}
  \hline
  \bf{R} &$E_{int}^{KS}$ & $E_{int}^{KS}$ & $\braket{\Delta\rho}$  \\
    (\AA)& (kcal/mol) & (kcal/mol)  &   \\
  \hline
4 & 1.83 & -1.01 & 0.016\\
5 & 0.48 & 0.23 & 0.002\\
6 & 0.39 & 0.38 &0.000\\
7 & 0.28 & 0.28 &0.000\\
8 & 0.19 & 0.19 &0.000\\
  \hline
 \end{tabular}
\end{table}
KS-DFT predicts a repulsive interaction energy for all distances, consistent with the known deficiency of semilocal GGA functionals to account for van der Waals interactions 
typical for such $\pi-\pi$ 
stacked complexes.\cite{goer2011} Subsystem DFT, on the other hand, predicts a negative interaction energy at the shortest considered intermolecular distance of $4$ \AA. Such 
behavior of subsystem DFT has been reported and analyzed in detail recently\cite{geno2014, kevo2006, kevo2014, beyh2013,schl2015} and while the negative interaction energy is more 
consistent with the physical 
interactions present in the system, it is most likely the result of error compensation of the NAKE which struggles with the present density overlap at this 
short intermolecular distance. For intermolecular distances of $6$ \AA\ and larger, the difference between DFT and subsystem DFT interaction energies becomes negligible. This is 
confirmed by the $\braket{\Delta\rho}$ values which show that for these distances, FDE reproduces the KS-DFT density. 

\begin{figure}[htp]
  \centering
  \caption{Optical spectra of the benzene-fulvene dimer at the separation distances of  $R=4$, $6$ and $8$ \AA, obtained using rt-TDDFT and subsystem rt-TDDFT methods.}
  \begin{subfigure}[b]{\textwidth}
     \includegraphics[scale=0.5]{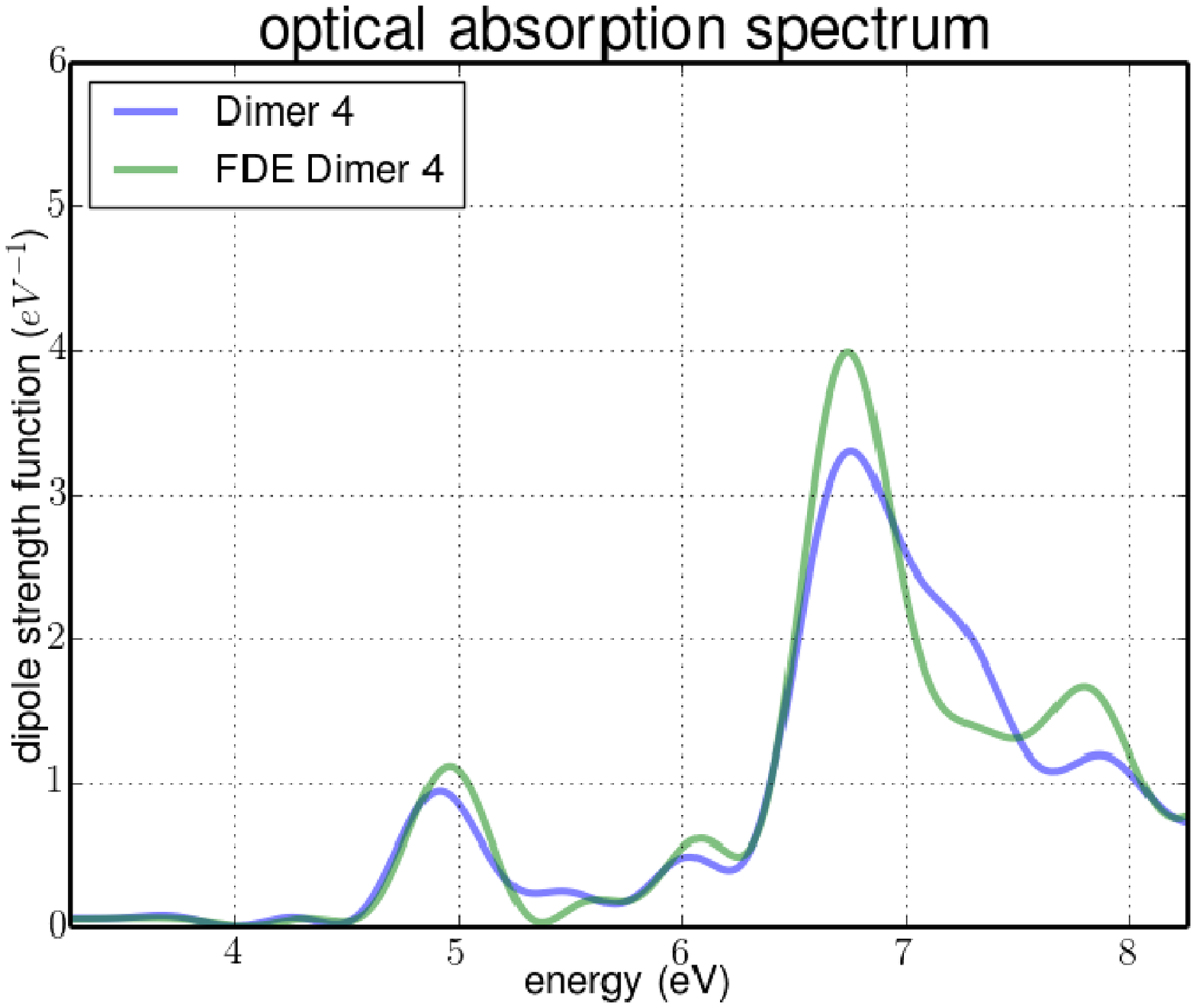}
     \caption{}
     \label{subfig:dimer4_fde}
  \end{subfigure}   
\begin{subfigure}[b]{\textwidth}
     \includegraphics[scale=0.5]{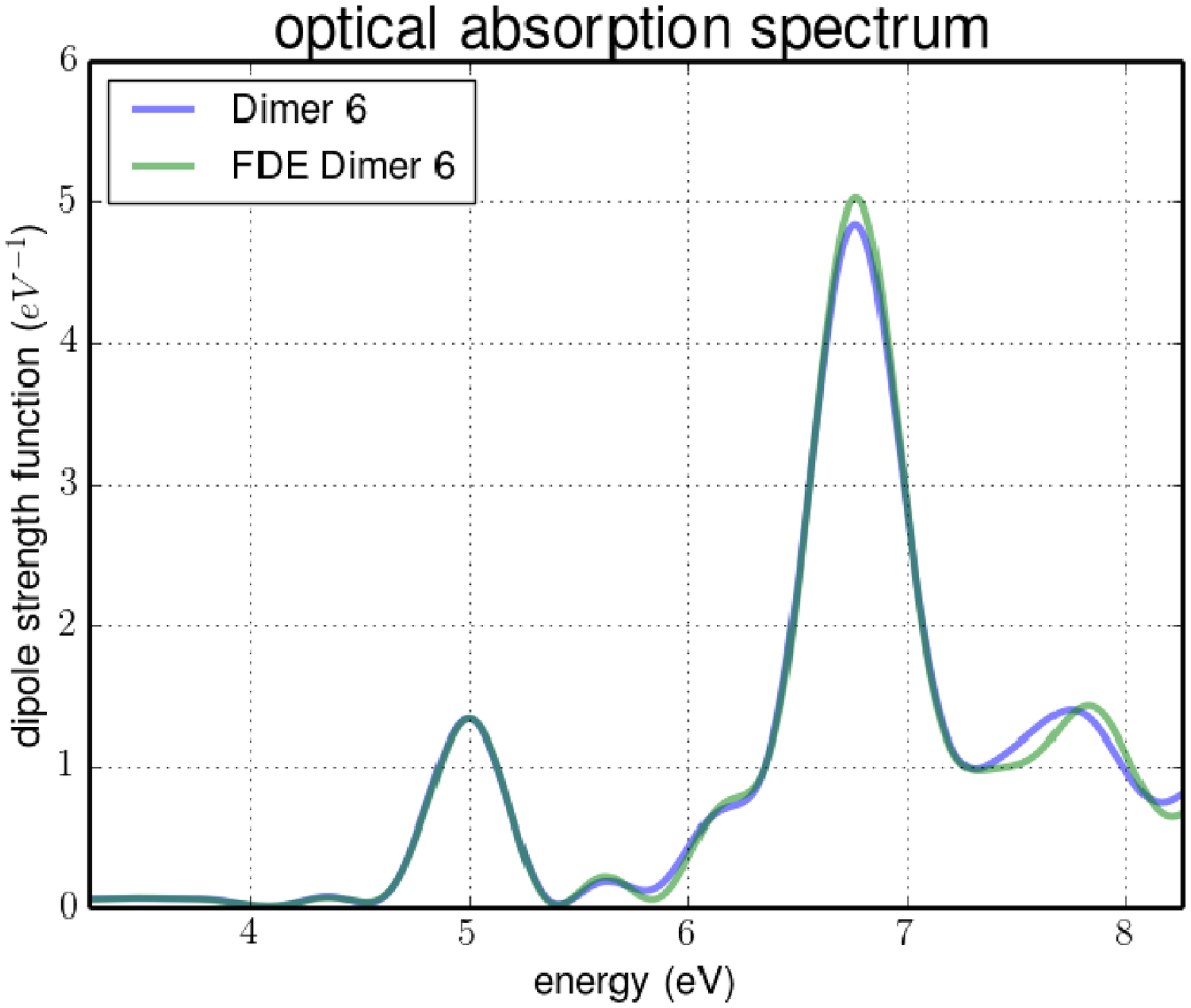}
     \caption{}
     \label{subfig:dimer6_fde}
  \end{subfigure}   
  \begin{subfigure}[b]{\textwidth}
     \includegraphics[scale=0.5]{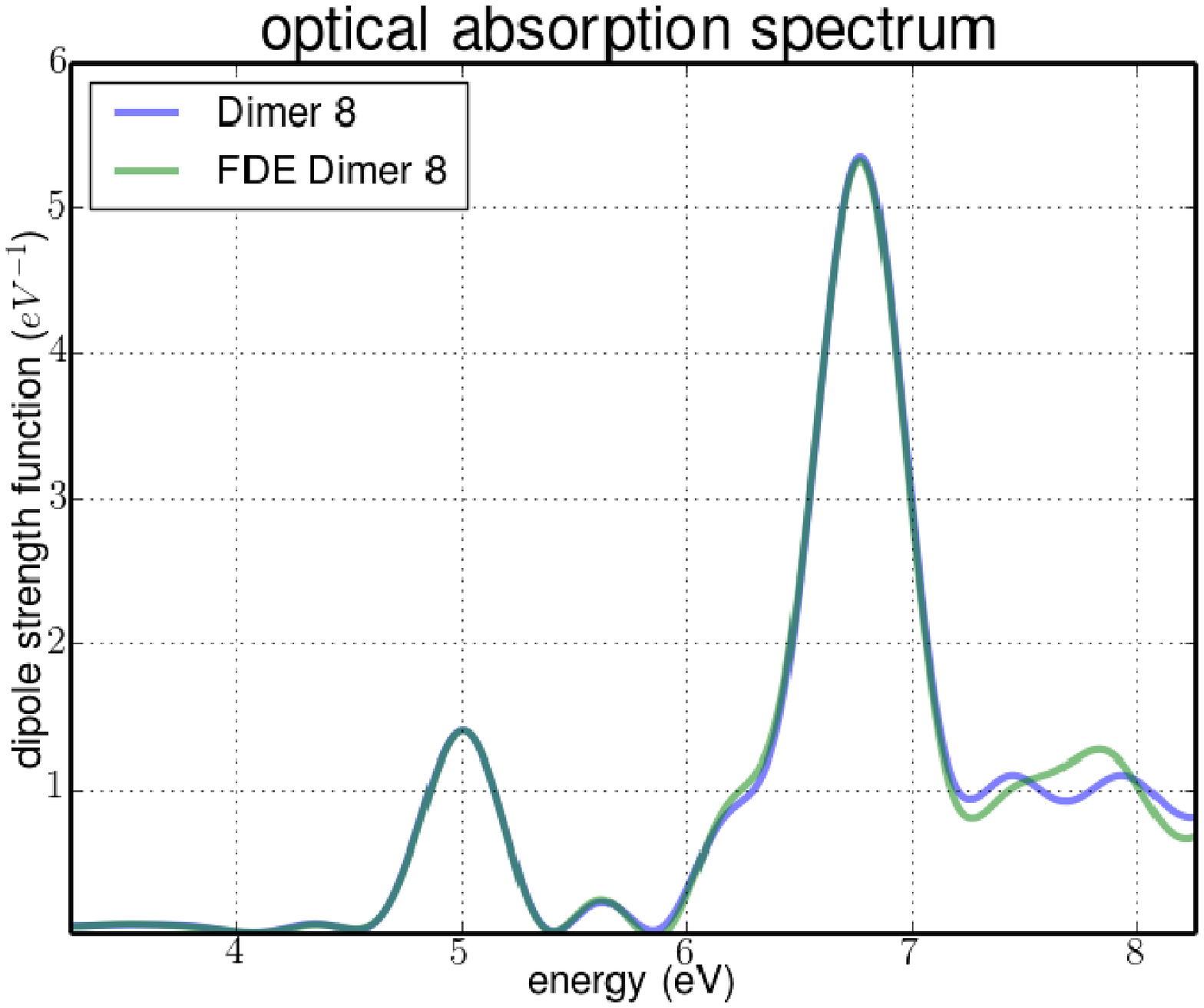}
     \caption{}
     \label{subfig:dimer8_fde}
  \end{subfigure}  
\label{fig:tot_fde_dimer}
\end{figure}

\begin{figure}[htp]
  \centering
  \caption{The time evolution of the dipole moment in the $x$ direction for the intermolecular distances of $6$ and $8$ \AA, obtained using rt-TDDFT and subsystem rt-TDDFT 
methods.}
  \begin{subfigure}[b]{\textwidth}
     \includegraphics[scale=0.5]{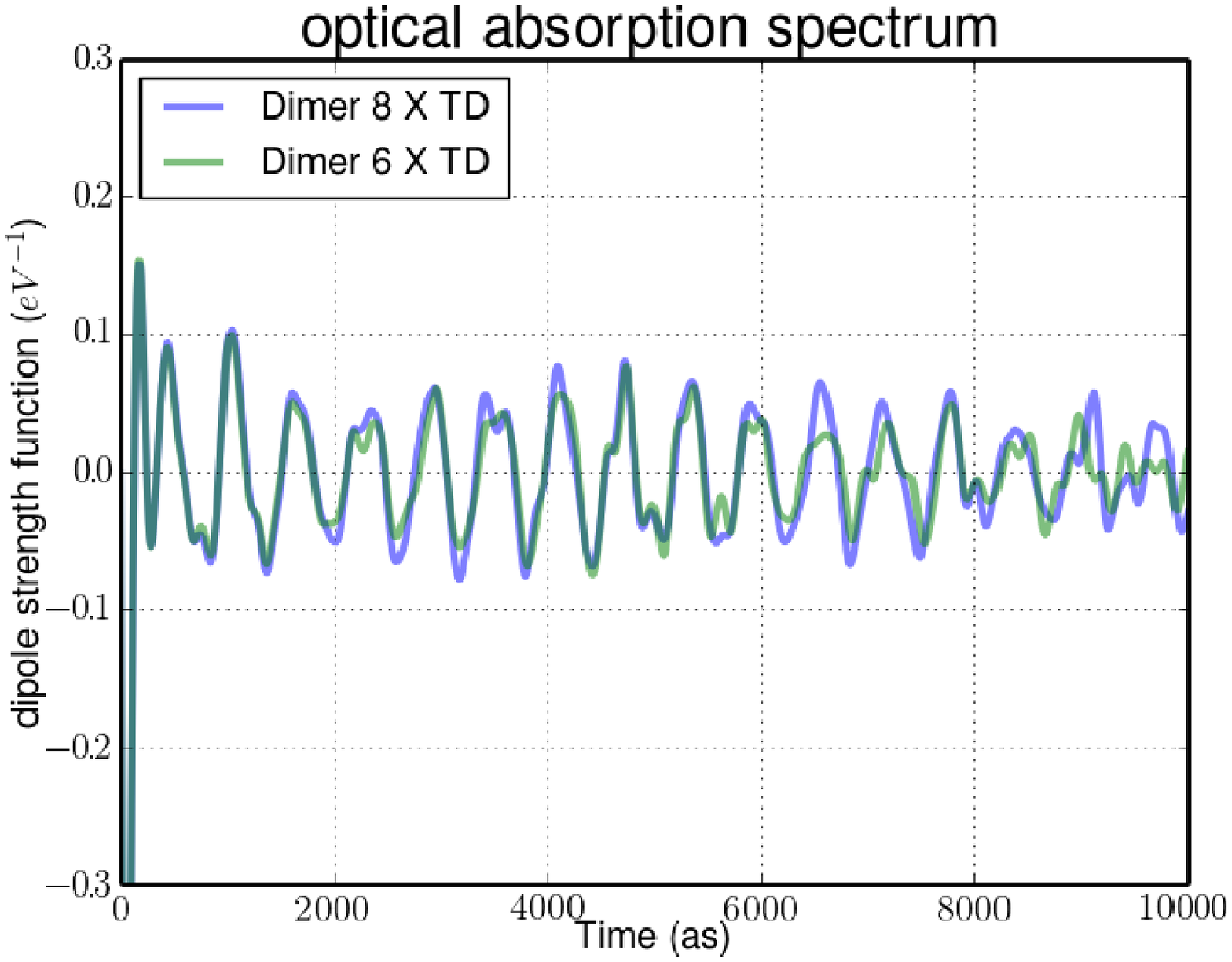}
     \caption{}
     \label{subfig:dimer6_8_td}
  \end{subfigure}   
\begin{subfigure}[b]{\textwidth}
     \includegraphics[scale=0.5]{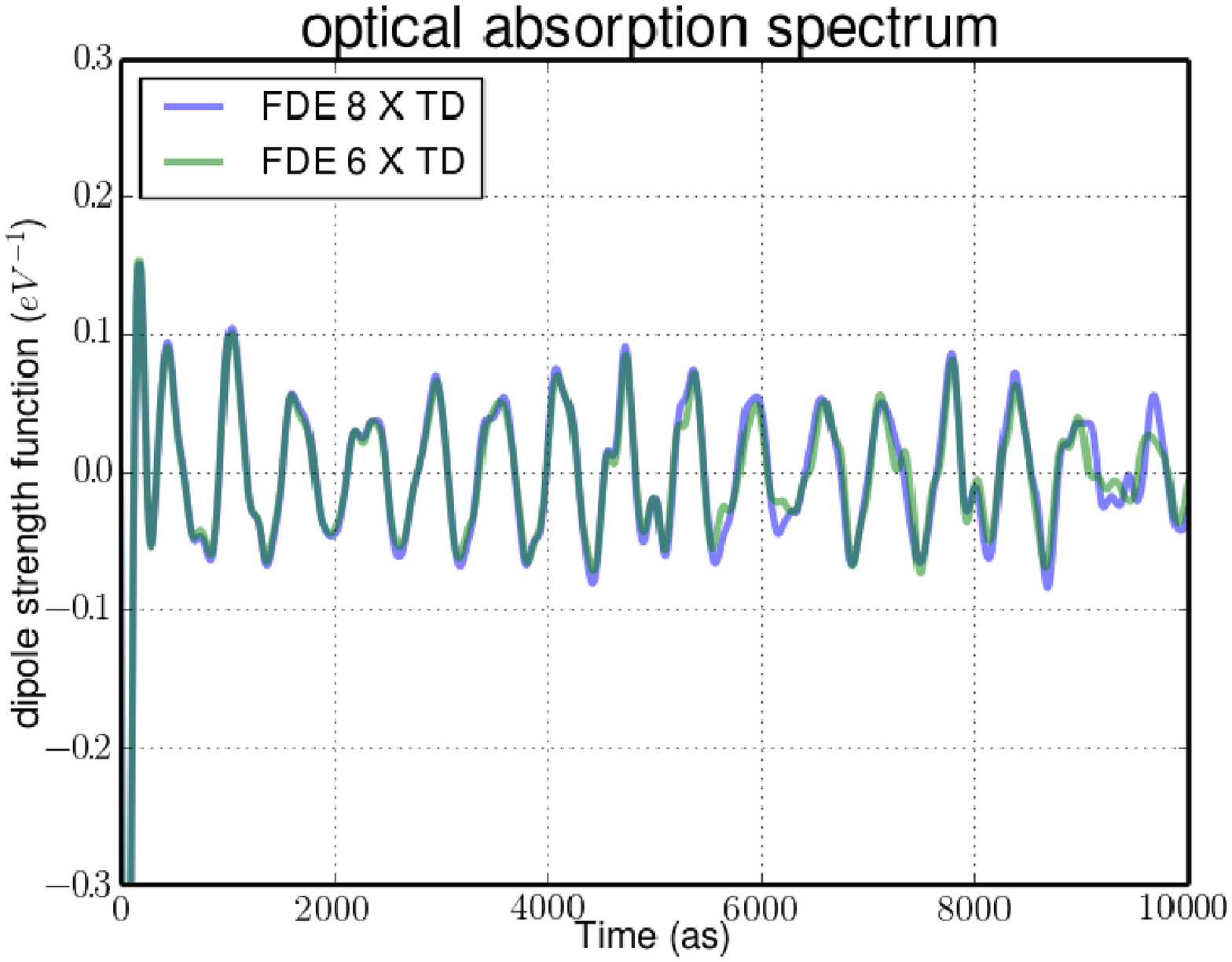}
     \caption{}
     \label{subfig:fde6_8_td}
  \end{subfigure}   
  \label{fig:dipole_td}
\end{figure}

The accuracy of the subsystem 
rt-TDDFT method, compared to supermolecular rt-TDDFT, is depicted in Fig.\ \ref{fig:tot_fde_dimer}, where the total optical absorption spectrum of the 
benzene-fulvene dimer obtained using the two methods is shown for the three  separation distances $R=4$, $6$ and $8$ \AA.
For $R=4$ \AA, the difference between the supermolecular and subsystem calculation is the largest, which is consistent with the $\braket{\Delta\rho}$ value in Table \ref{tab:eint}  
of the dimer 
from the ground state calculations and can be attributed to the failure of the nonadditive kinetic energy functionals.\cite{neug2007} For the intermolecular distance of $R=6$ \AA, 
both methods reproduce very similar spectra, with the supermolecular calculation showing three major excitations at $4.99$ , $6.78$  and $7.73$ eV and subsystem calculation having 
very similar values of $4.99$, $6.77$ and  $7.78$ eV. These deviations are well within the accuracy of $0.06$ eV established for the FDE formulation of the LR-TDDFT 
method.\cite{weso2002b} The oscillator strength at the frequency of the lowest/highest excitation energies are identical to KS-DFT and a slight difference is 
found for 
the middle excitation energy. Surprisingly, the situation becomes worse for the largest intermolecular distance of $R=8$ \AA, where the spectrum is completely identical 
below $7.2$ eV but diverges for the higher frequencies, where the KS-DFT calculation produces a flattened spectrum but FDE retains the peak at $7.73$ eV as in the spectrum of 
$R=6$ 
\AA. In fact, the spectra obtained for the two intermolecular distances using subsystem 
DFT are quasi identical, suggesting that FDE overestimates the intermolecular interaction at the larger separation distance compared to supermolecular rt-TDDFT. This effect can 
also be seen from the time evolution of the dipole moment in the subsystem and supermolecular calculations, which are depicted for the two intermolecular distances in Fig.\ 
\ref{fig:dipole_td} for the first $10$ femtoseconds . Only the $x$ component of the dipole moment is shown, from which the peak in question originates. As one can see, there is a 
discernible difference in the evolution of the dipole moment for the two distances in the supermolecular calculation, while in the subsystem calculation the difference between the 
two distances is very small and both resemble the evolution of the dipole moment in the supermolecular calculation for $R=6$ \AA. Further research is required to determine the 
physics behind this phenomenon. 

\begin{figure}[htp]
  \centering
  \caption{Optical spectra of the  isolated benzene and fulvene molecules, obtained using rt-TDDFT, compared to the spectra of the embedded molecules in the dimers at the 
separation distances of  $R=4$, $6$ and $8$ \AA., obtained using subsystem rt-TDDFT.}
  \begin{subfigure}[b]{\textwidth}
     \includegraphics[scale=0.5]{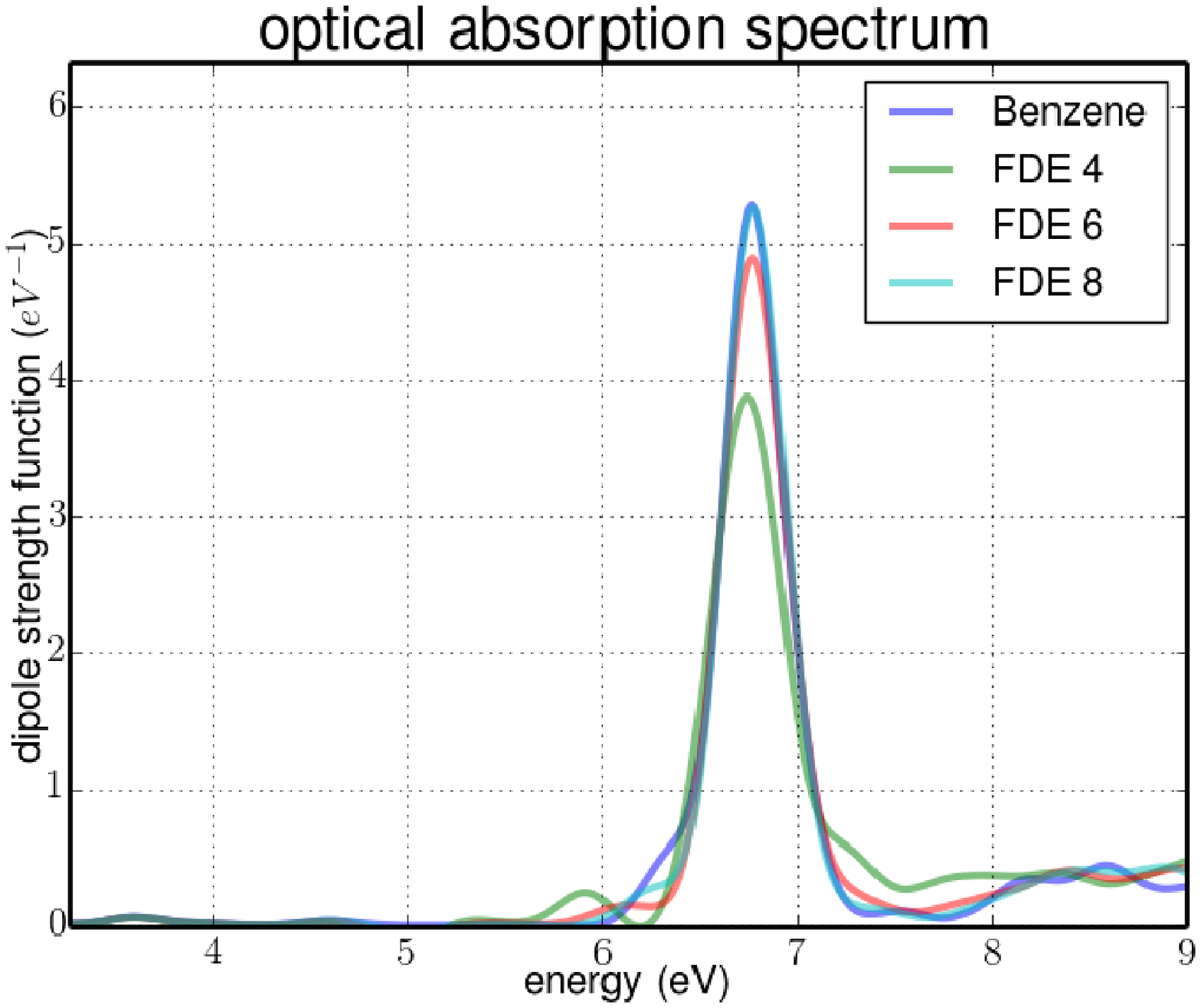}
     \caption{}
     \label{subfig:mono1_fde40}
  \end{subfigure}   
\begin{subfigure}[b]{\textwidth}
     \includegraphics[scale=0.5]{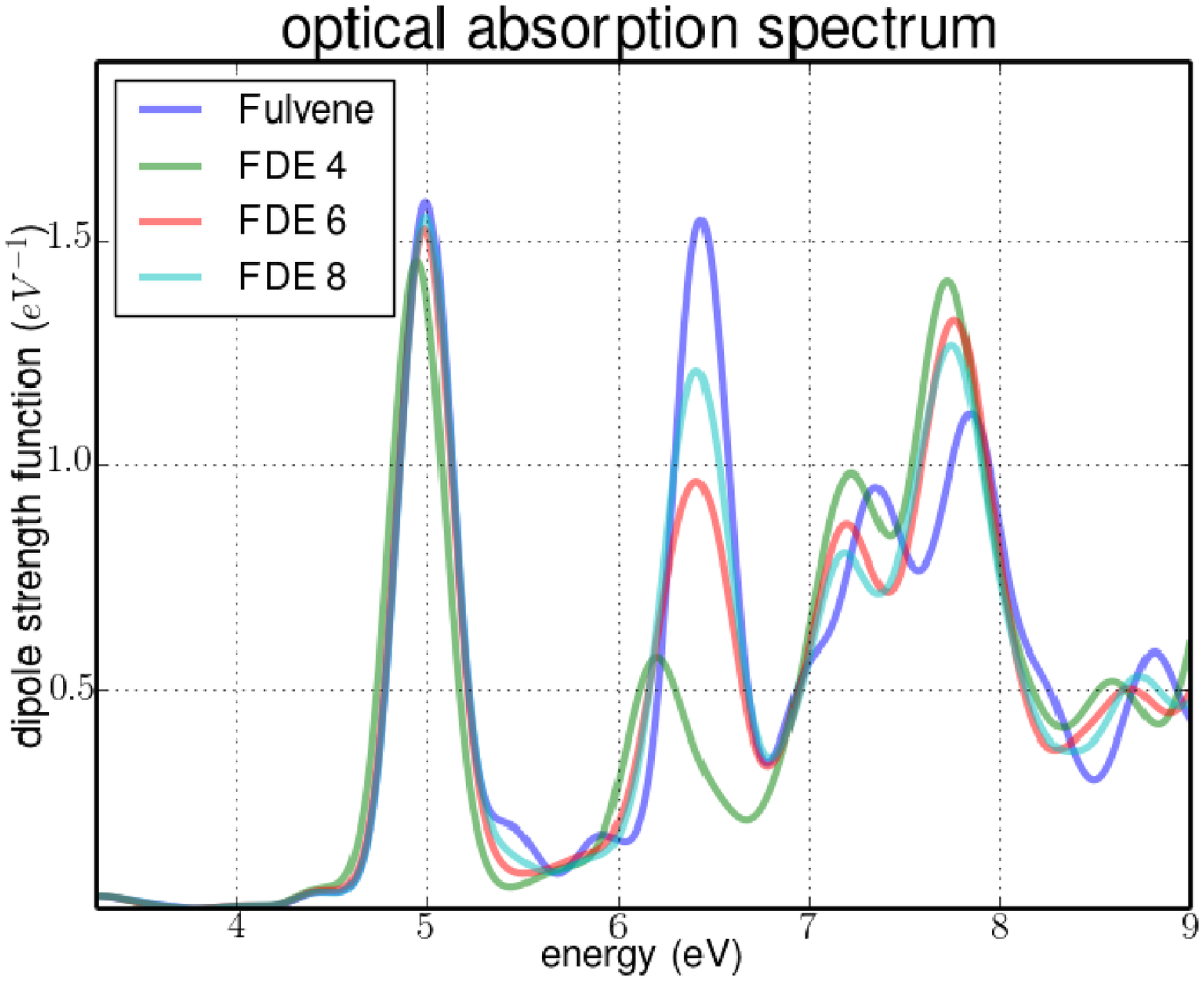}
     \caption{}
     \label{subfig:mono2_fde41}
  \end{subfigure}   
  \label{fig:mono_fde_468}
\end{figure}

Additional to reproducing the supermolecular rt-TDDFT result in a divide-and-conquer manner, the subsystem approach has the interesting feature of allowing to compare properties 
of isolated systems with embedded systems, providing first hand information on the effect of the embedding on the properties of the subsystems. Fig.\ \ref{fig:mono_fde_468} shows 
the optical spectra of the isolated benzene and fulvene molecules, compared to their embedded equivalents in the dimers at different intermolecular 
separations. The benzene molecule only shows one peak at $6.77$ eV for the isolated molecule. At $R=8$ \AA, the subsystem calculation reproduces the supermolecular result exactly, 
at $R=6$ \AA\ the position of the peak is correct but the oscillation strength is slightly underestimated and at $R=4$ \AA\ there is also a slight interaction induced shift to the 
lower frequency of $6.75$ eV. The optical spectrum of the fulvene molecule is more complicated due to lower symmetry, with four major excitations at $4.99$, $6.44$, $7.30$ and 
$7.90$ eV. The first three excitations originate from the dipole response to an electric field in the $x$ direction, while the last one stems from a response to an electric field 
in the $y$-direction. The optical spectra from the subsystem calculation show a very clear effect from the embedding potential generated by the presence of the benzene molecule 
even at the separation distance of $8$ \AA. For the lowest two peaks and the highest peaks, the subsystem spectra are shifted to lower energies for 
the smallest intermolecular distance of $R=4$. The shift is smallest for the excitation energy of $4.99$ eV  ($0.05$ eV) and largest for the excitation energy of $6.44$ ($0.17$). 
For 
the intermolecular distances of $R=6$ and $8$, the correct excitation energy is reproduced and the effect of the embedding is visible only in the lower oscillator strength, which 
slowly converges to the value in the isolated molecule. At the highest excitation energy of $7.90$ eV, there is also an interaction induced shift of $0.13$ eV at 
the $R=6$ and $8$ \AA\  and the intensity is overestimated for the shorter distances. The situation for the peak at $7.30$ eV is different, where all 
embedded fulvene spectra show similar shift of $0.13$ eV ($0.14$ eV for $R=4$). Additionally, the oscillator strength diverges from the isolated value with increasing 
intermolecular distance.  

The question of the interaction induced shift has been previous discussed within FDE using the linear response formulation of the 
method.\cite{neug2005d,neug2005e,jaco2006,weso2004,neug2006b} In the original implementation of linear response FDE,\cite{casi2004} the linear density 
response of a subsystem is obtained while keeping the densities of the other subsystems frozen in time. As a result, the response is restricted to the active subsystem 
only.\cite{casi2004,weso2004} As mentioned above, this uncoupled response, $\chi^{u}_I$, is equivalent to performing a subsystem rt-TDDFT calculation where only the KS orbitals of 
one subsystems are 
propagated in time, while the other subsystems are kept in their initial state. The advantage of such an approach is the considerable computational simplicity. 
Numerous applications on optical spectra\cite{neug2005e,neug2005b}, Raman spectra\cite{neug2005c} induced circular dichroism\cite{neug2005d,neug2006b} and electron-spin-resonance 
hyperfine 
couplings\cite{neug2005f} have shown that the uncoupled response is sufficient in reproducing supermolecular results with good accuracy, even in the presence of hydrogen bonds, as 
long as there are no couplings in the excitation between the systems. An example of when the uncoupled response is not enough is the case of identical chromophores, where explicit 
excitation transfer occurs, as will be discussed in Section \ref{eet}. In later work, J.\ Neugebauer\cite{neug2007} introduced a framework for including coupling between 
selected excitations in the linear response regime. The subsystem rt-TDDFT method offers a straightforward way to include couplings between \textit{all} excitations, as well as 
reproducing the uncoupled 
results. Clearly, a fully coupled calculation is more computationally expensive than an uncoupled one.

\begin{figure}[htp]
  \centering
  \caption{Optical spectra of the embedded benzene molecule, obtained using coupled and uncoupled subsystem rt-TDDFT methods at the 
separation distances of  $R=8$, $6$ and $4$ \AA.}
  \begin{subfigure}[b]{\textwidth}
     \includegraphics[scale=0.5]{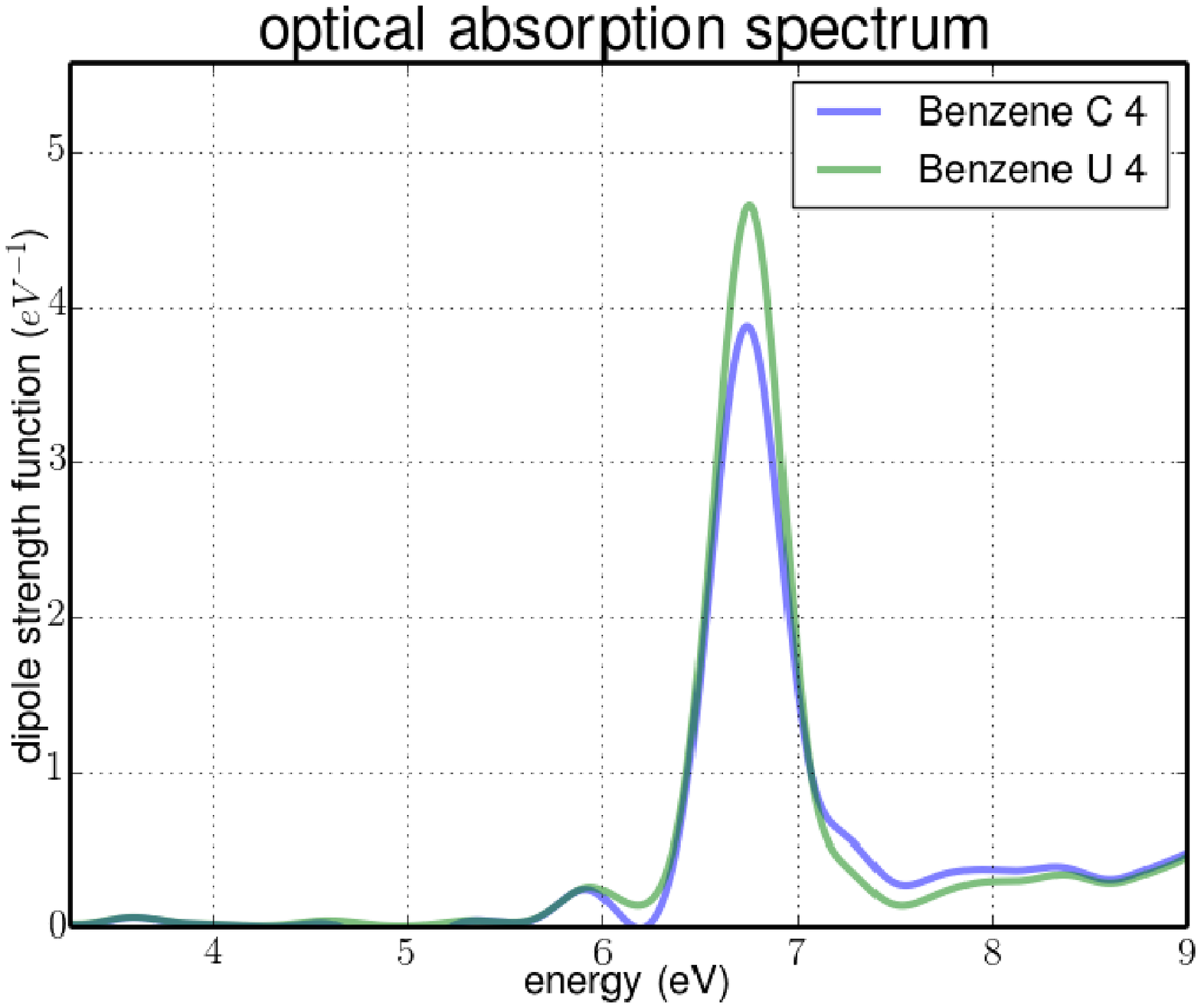}
     \caption{}
     \label{subfig:fde40_eg4fde0}
  \end{subfigure}   
\begin{subfigure}[b]{\textwidth}
     \includegraphics[scale=0.5]{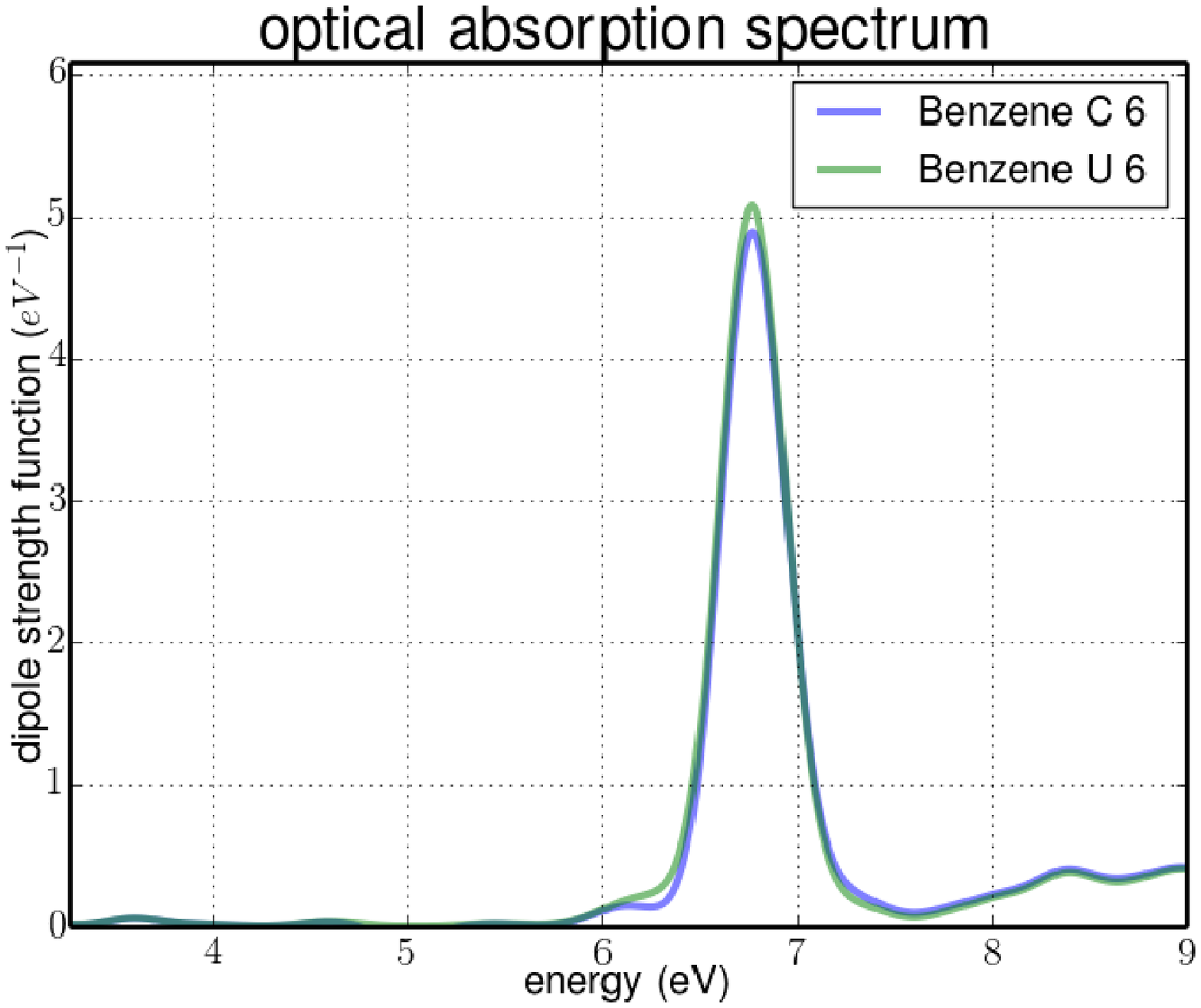}
     \caption{}
     \label{subfig:fde60_eg6fde0}
  \end{subfigure}   
  \begin{subfigure}[b]{\textwidth}
     \includegraphics[scale=0.5]{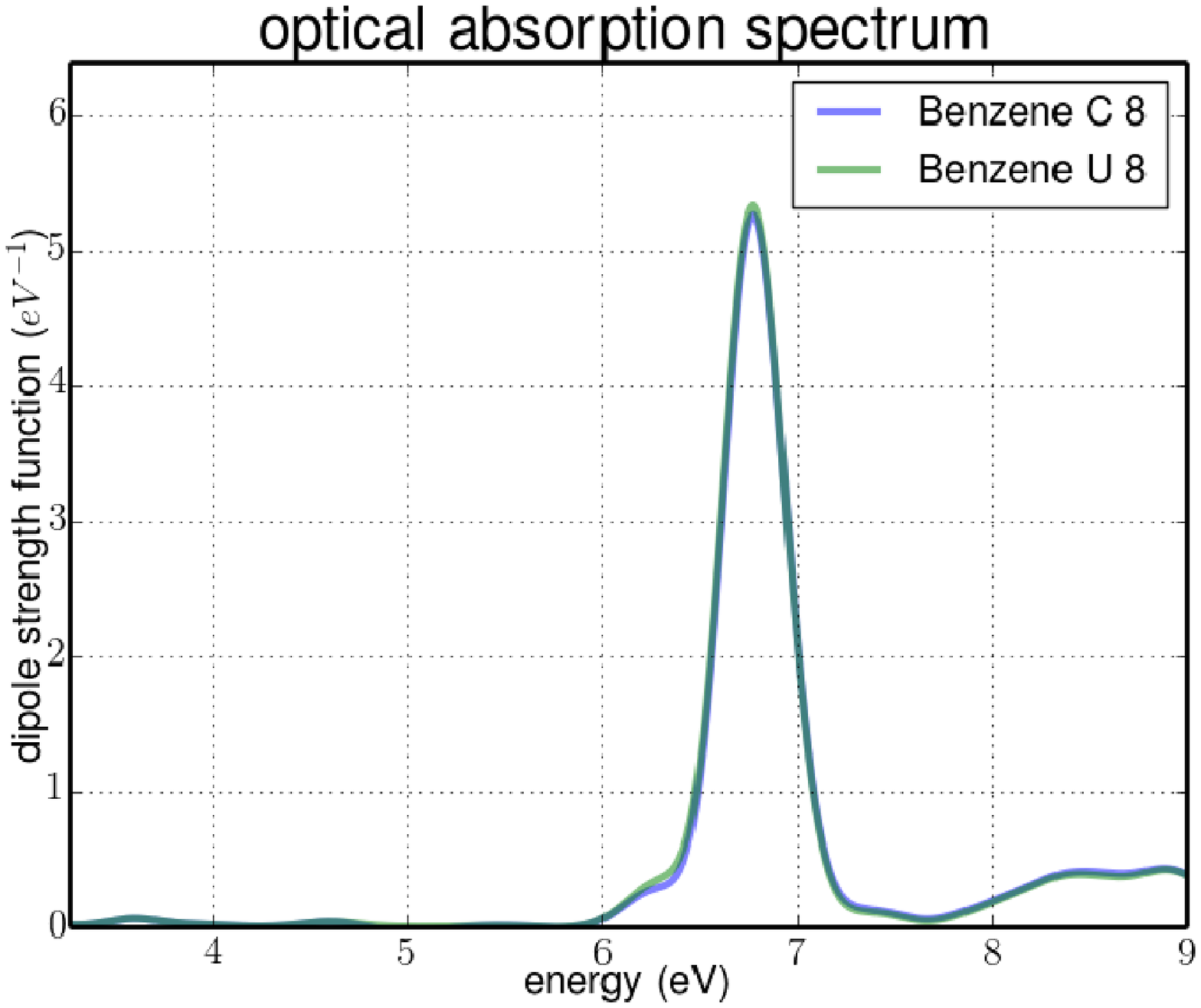}
     \caption{}
     \label{subfig:fde80_eg8fde0}
  \end{subfigure} 
  \label{fig:BenzeneCU}
\end{figure}

\begin{figure}[htp]
  \centering
  \caption{Optical spectra of the embedded fulvene molecule, obtained using coupled and uncoupled subsystem rt-TDDFT methods at the 
separation distances of  $R=8$, $6$ and $4$ \AA.}
  \begin{subfigure}[b]{\textwidth}
     \includegraphics[scale=0.5]{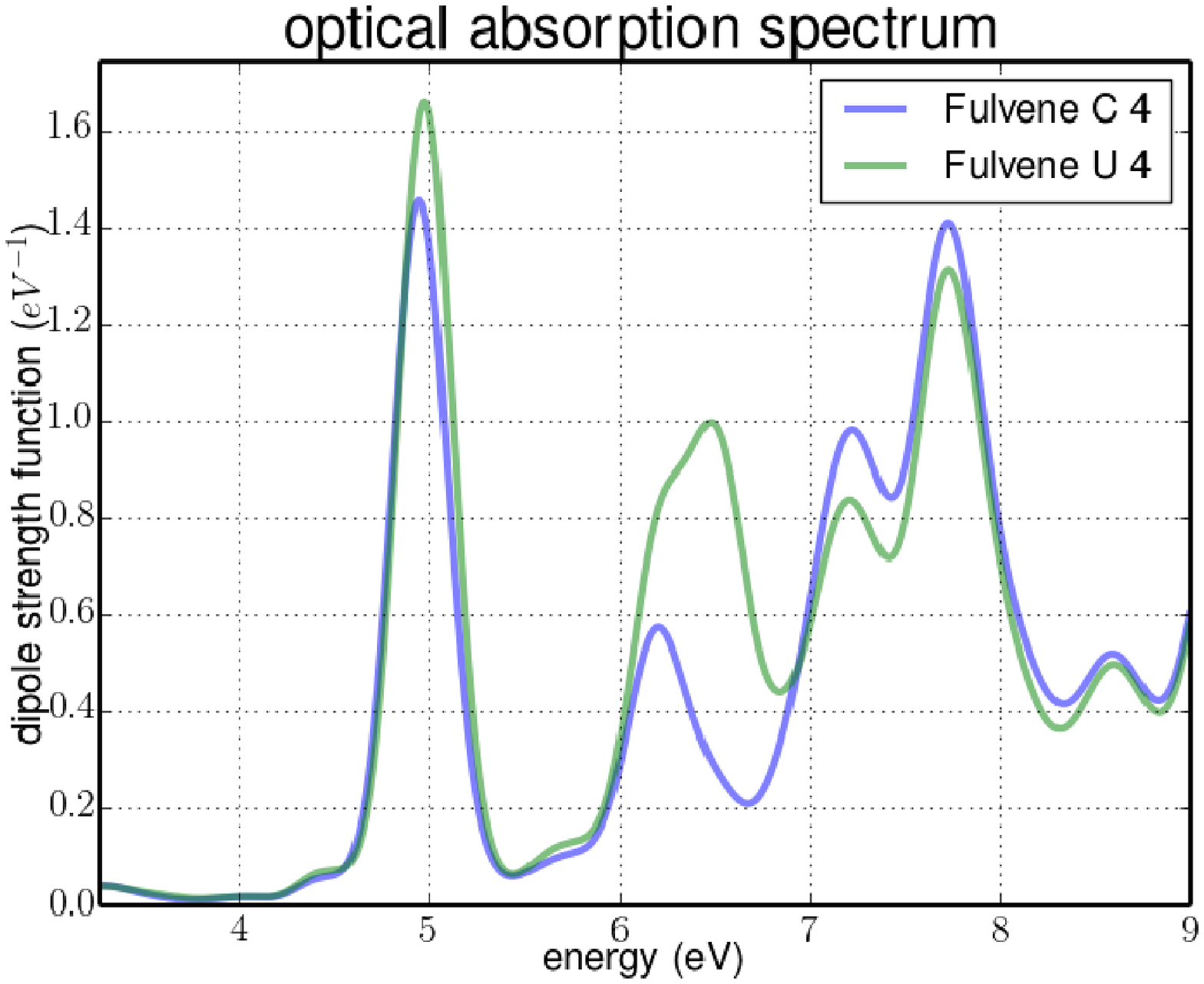}
     \caption{}
     \label{subfig:fde41_ge4fde1}
  \end{subfigure}   
\begin{subfigure}[b]{\textwidth}
     \includegraphics[scale=0.5]{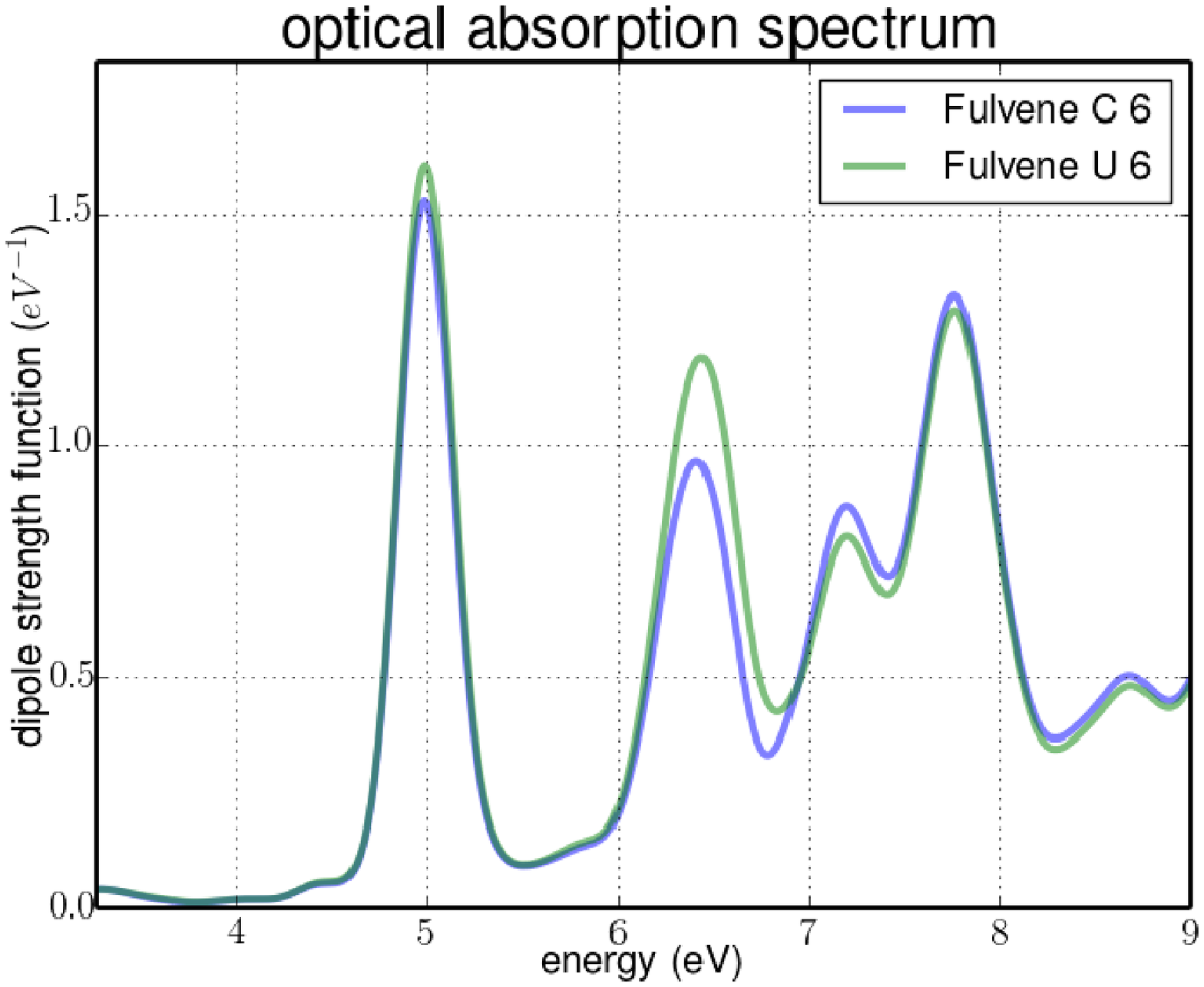}
     \caption{}
     \label{subfig:fde61_ge6fde1}
  \end{subfigure}   
  \begin{subfigure}[b]{\textwidth}
     \includegraphics[scale=0.5]{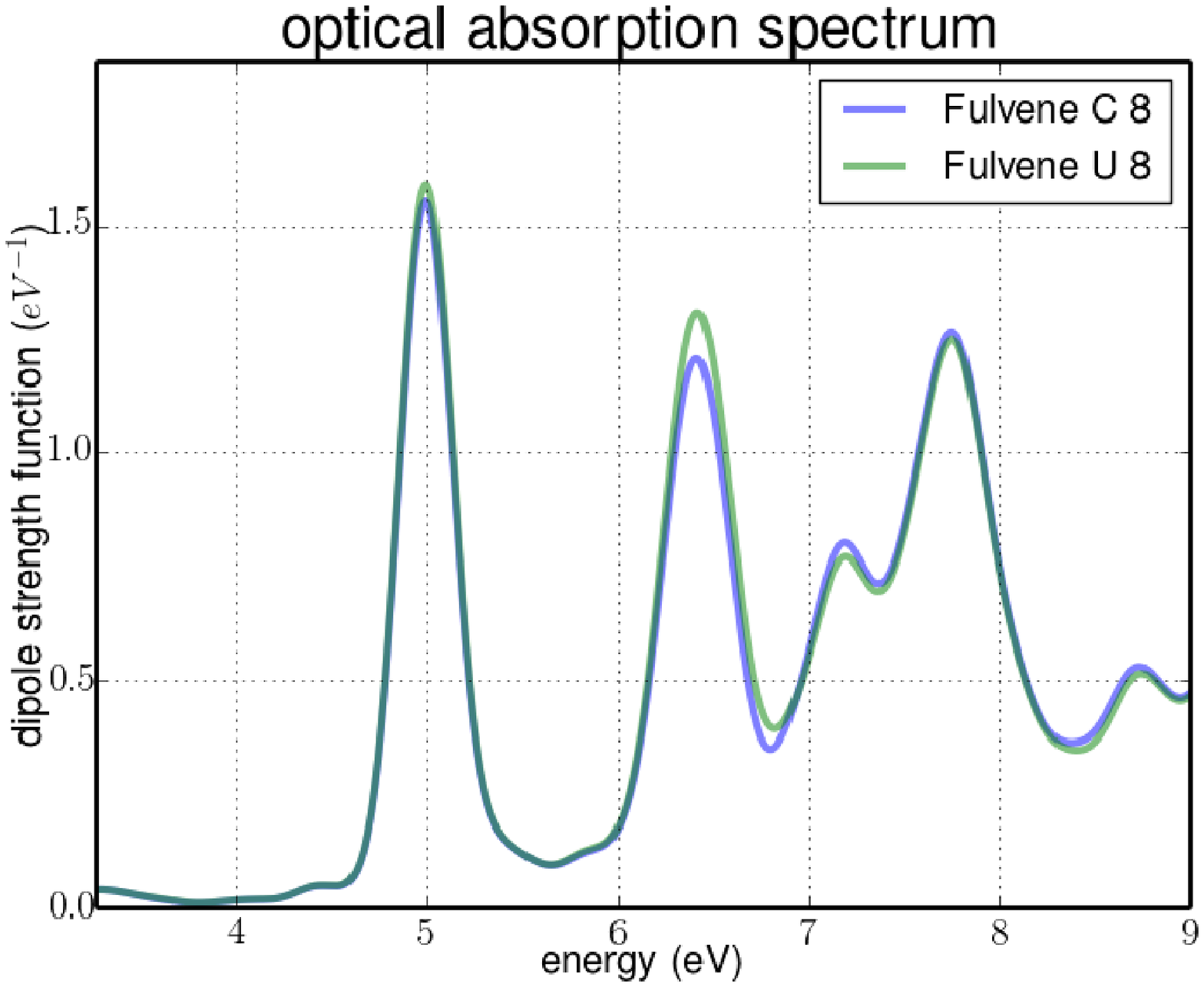}
     \caption{}
     \label{subfig:fde81_ge8fde1}
  \end{subfigure} 
  \label{fig:FulveneCU}
\end{figure}

The difference in the results between the coupled and uncoupled calculations for the benzene-fulvene dimer is 
shown in Figs.\ \ref{fig:BenzeneCU} and \ref{fig:FulveneCU} for different intermolecular separations.
The largest differences between the coupled and uncoupled results are found for the fulvene molecule for the excitation energy at $6.44$ eV for both the interaction induced 
shift and the oscillation strength. In the fully coupled calculation, the dynamical embedding potential due to the presence of the benzene molecule results in a shift of $0.20$ eV 
at the 
separation distance of $4$ \AA, while no shift is present for the uncoupled calculation (i.e.\ stationary embedding). Additionally, a large difference in the oscillation strength 
for this excitation is found 
between the coupled and uncoupled calculations. Though not as pronounced, a similar effect is found for the $4.99$ eV excitation in fulvene and $6.78$ eV in 
benzene. At $R=4$ \AA, the coupled calculation produces an interaction induced shift of $0.04$ eV, compared to $0.02$ eV in the uncoupled 
calculation, and a lower oscillation strength. At the intermediate distance of $6$ \AA, the effect of the lower oscillation strength is still present, but no interaction induced 
shift is observed. At $8$ \AA, the coupled and uncoupled spectra are almost identical, with the exception of the $6.44$ eV excitation in the fulvene molecule, where there is still 
a small effect on the oscillation strength. It should also be noted that for the uncoupled excitations in the fulvene spectrum, the coupled calculation produces slightly 
larger oscillation strengths, which is necessary in order to satisfy the sum rule of the spectrum. 

\subsection{Excitation energy transfer}
\label{eet}
\begin{figure}[htp]
 \centering
 \caption{Applied electric field to the Na$_4$ cluster.}
  \label{fig:ef}
 \includegraphics[scale=0.5]{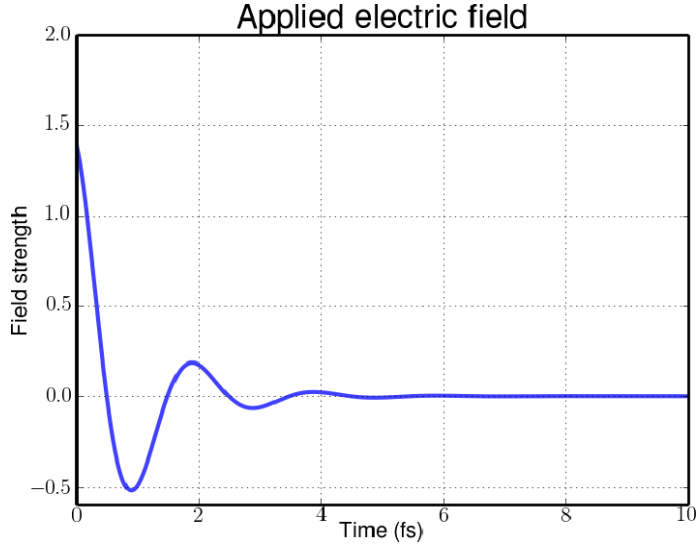}
\end{figure}

\begin{figure}[htp]
  \centering
  \caption{Excitation energy transfer between two Na$_2$ molecules at the separation distances of $12.5$, $15$ and $17.5$ Bohr, after the excitation of the donor molecule with an 
applied electric field of $2.18$ eV frequency.}
  \begin{subfigure}[b]{\textwidth}
     \includegraphics[scale=0.5]{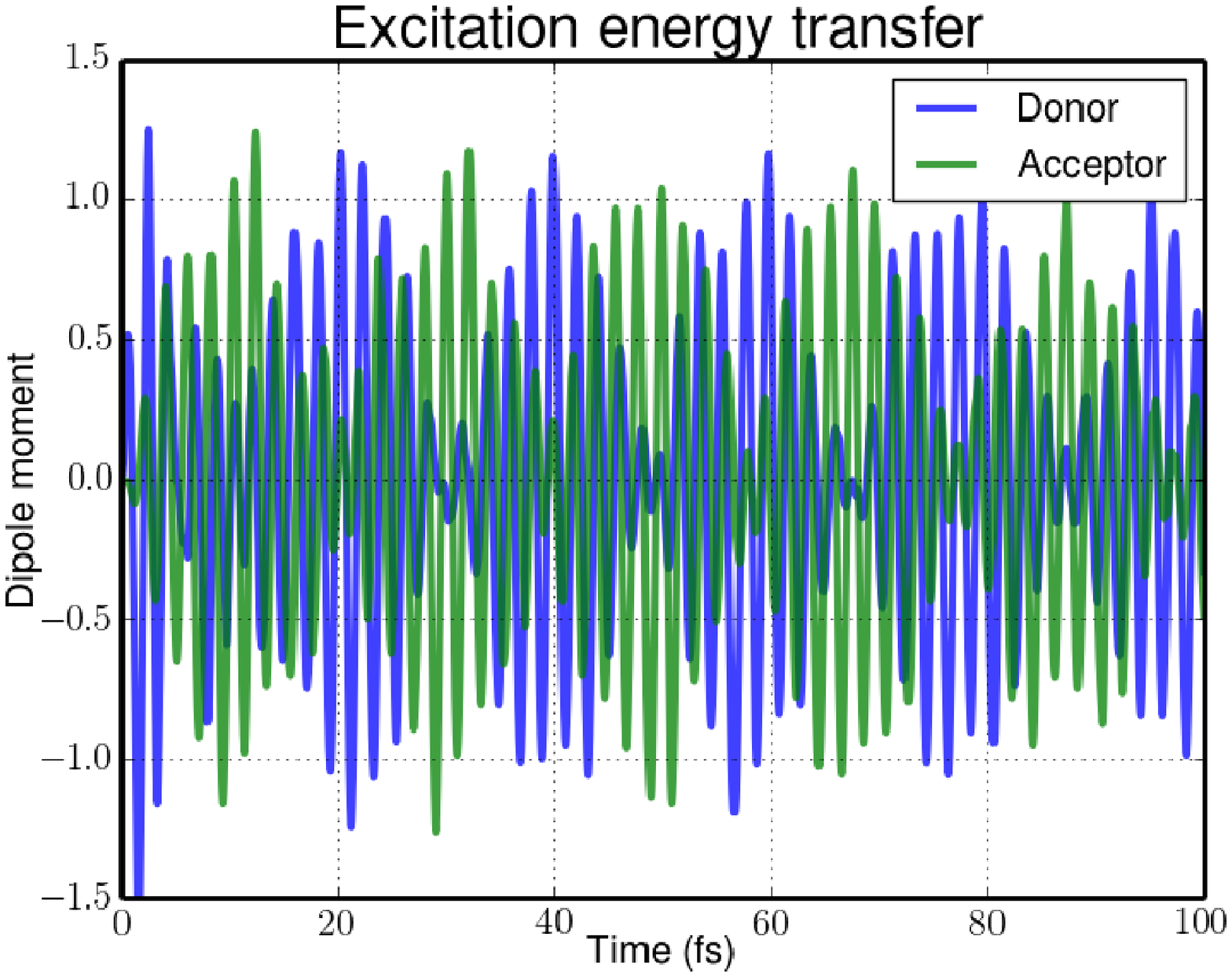}
     \caption{}
     \label{subfig:12_cos3}
  \end{subfigure}   
\begin{subfigure}[b]{\textwidth}
     \includegraphics[scale=0.5]{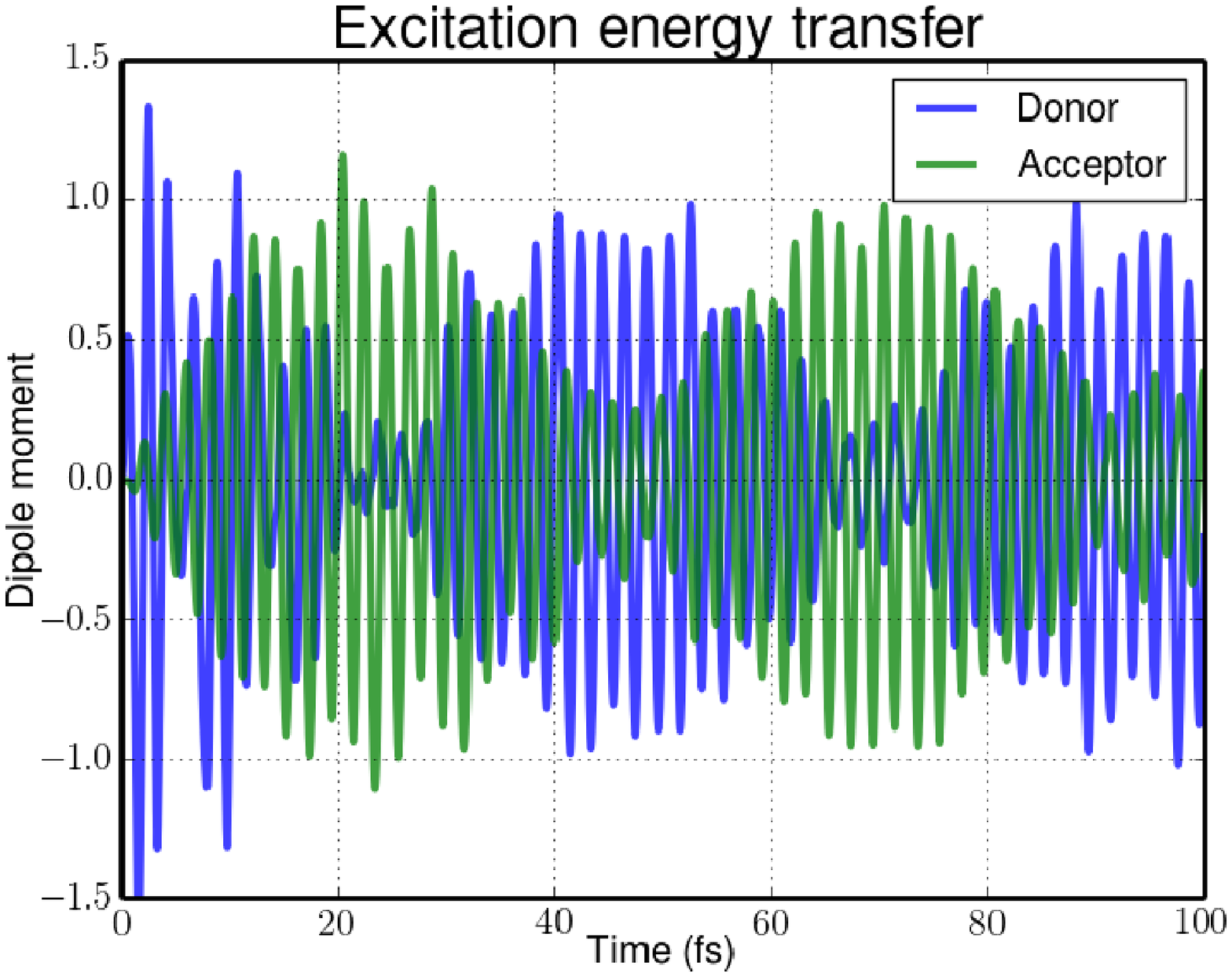}
     \caption{}
     \label{subfig:15_cos3}
  \end{subfigure}   
  \begin{subfigure}[b]{\textwidth}
     \includegraphics[scale=0.5]{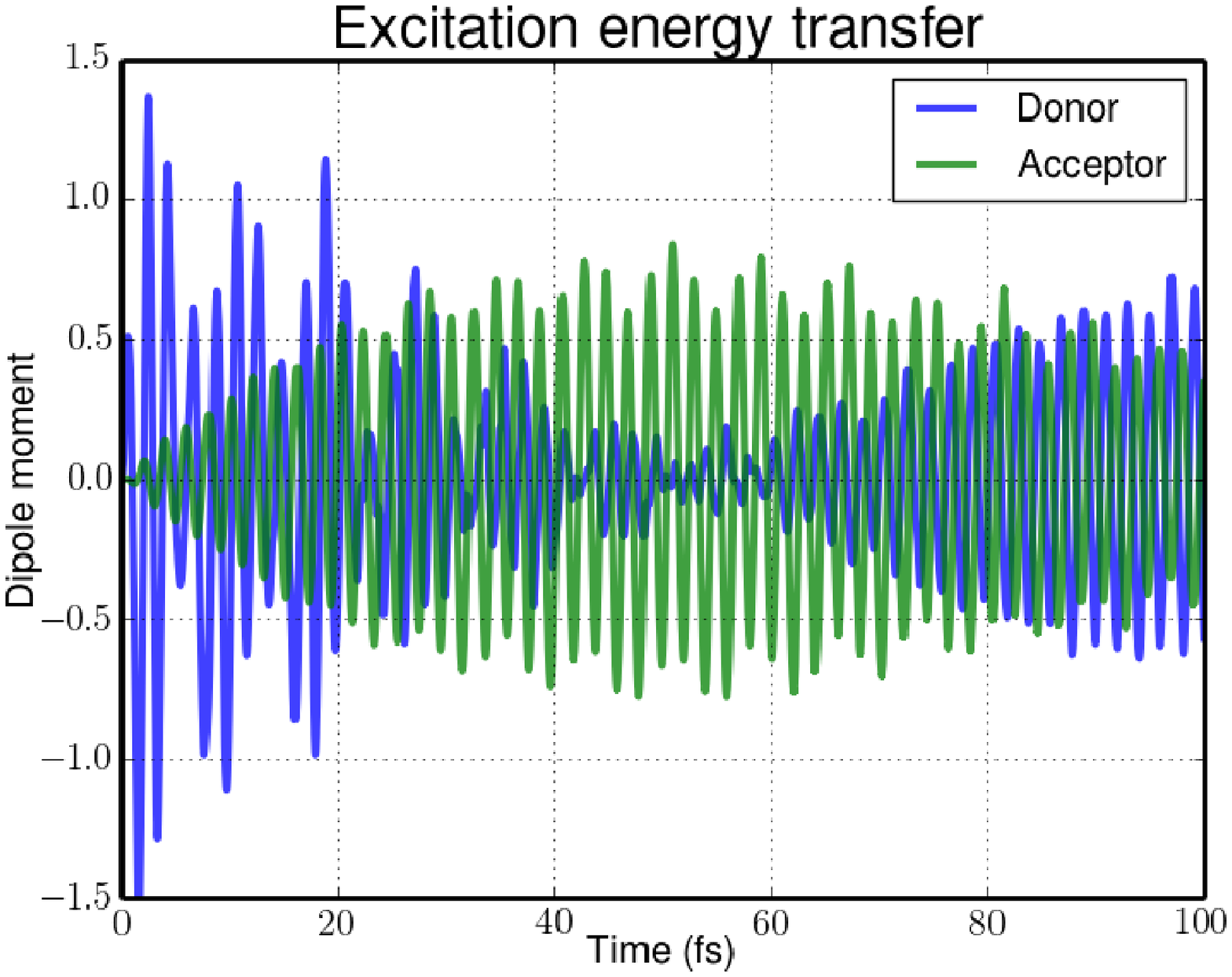}
     \caption{}
     \label{subfig:17_cos3}
  \end{subfigure} 
  \label{fig:na4}
\end{figure}

In the last section, we have seen that subsystem rt-TDDFT allows us to reproduce full coupling between subsystems in a straightforward fashion. A 
strongly coupled system which is often studied using TDDFT is a dimer consisting of two identical chromophores. An adequate description of such system requires a method which 
includes the full coupled response between all subsystem\cite{neug2007}, as clearly shown by Eq.\ (\ref{eq:dyson}).\cite{pava2013b} The coupling in excitations along a certain 
direction between the chromophores opens a window for an explicit excitation energy transfer (EET). The central quantity in EET is the electronic 
coupling $V$, defined as
\begin{equation}
 V=\braket{D^*A|\hat{H}|DA^*}
\end{equation}
where $\ket{D^*A}$ and $\ket{DA^*}$ are orthogonal diabatic states.
Herein, $D$ and $A$ represent the donor and acceptor in the excitation energy exchange transfer process and $\hat{H}$ is the molecular Hamiltonian 
operator.\cite{foer1948,foer1965,dext1953}. The coupling is linked to the excitation energy transfer rate $k^{ET}$ through Fermi's Golden Rule\cite{scho2003}
\begin{equation}
 k^{ET}=2\pi |V|^2\int J(\epsilon)d\epsilon
\end{equation}
where $J(\epsilon)$ is the spectral overlap between the absorption spectra of the $D$ and $A$ chromophores. Within the framework of linear response TDDFT, the electronic coupling 
can be obtained from the Davydov 
splitting\cite{davydov_excitons,sagv2009} of the coupled excitation energy in the spectrum, compared to the absorption spectrum of the monomer. 
Real time TDDFT, however, offers the advantage to study such transport events in real time and this can offer an intuitive picture of the EET process. For this purpose, we chose 
the test system of two sodium dimers, situated along the 
$z$-axis with separation distances of $12$, $15$ and $17.5$ Bohr. This system has been well studied before\cite{hofm2010,huang2014,vasi2002} and is known to couple strongly at the 
excitation 
energy of $2.18$ eV in the direction of the Na-Na bond. The full cluster was placed in a supercell of $22.7 \times  22.7 \times 43.5 $ a.u.$^{3}$ to avoid periodic 
interactions. The calculation was performed with ultrasoft pseudopotentials from the 
GBRV\cite{garri2014} library with a kinetic energy cutoff of $55.0$ Ry and density cutoff of $660.0$ Ry and the LC94\cite{LC94} functional was used for the nonadditive kinetic 
energy. The exact value of the excitation energy was first obtained from an optical absorption spectrum of the Na$_4$ cluster using FDE. At the start of the 
simulation, the $\ket{D^*A}$ state is prepared by applying an electric field in a direction along the 
Na-Na bond to only one of the subsystems, noted as the donor, with a frequency of $2.18$ eV, using a cosine function damped by a Gaussian. This state is not stationary and will 
evolve accordingly by populating the $\ket{DA^*}$ state. The time evolution of the 
field is depicted in Fig.\ \ref{fig:ef}. Both of the subsystems are evolved simultaneously for $100$ fs with a time step of $2$ as. At each time step, the dipole moment in the 
direction of the excitation is recorded for both the donor and acceptor molecule, which are depicted in Fig.\ \ref{fig:na4} for the three separation distances.

Starting from $t=0$, the dipole moment of the donor molecule reacts to the applied field and the dipole moment starts oscillating, reaching a maximum within the $2.5$
fs, at which point the $\ket{D^*A}$ state reaches maximum population. From that point on, the system proceeds in transferring the excitation energy acquired from the 
applied field to the $\ket{DA^*}$ state. It is important to note that the acceptor subsystem has no direct field applied in its potential and the excitation 
transfer occurs solely through the response of the embedding potential to the perturbation. At the shortest separation distance of $12$ Bohr, the 
excitation energy is fully transfered to the $\ket{DA^*}$ state within approximately $20$ fs, $13$ fs after the damping of the applied field has taken place. The complete 
transfer of the excitation energy is clear from the intensity of the dipole moment when it reaches its maximum values, which is equal to the maximum  in the population of 
$\ket{D^*A}$ and $\ket{DA^*}$ states. As no nuclear relaxation is included in this simulation, the excitation energy 
is transfered back to the $\ket{D^*A}$ state. As can be seen from Fig.\ \ref{fig:na4}, the 
excitation rate strongly depends on the separation distance. According to F\"orster theory,\cite{foer1948,foer1965} the excitation transfer rate has a $R^n$ dependence 
with $n=6$ at long range and a lower $n$ dependence for shorter distances due to exchange interactions, as described by Dexter theory.\cite{dext1953} This has been previously 
confirmed for this particular system by Hofman et al.\cite{hofm2010}, which found the F\"orster theory to be valid for separation distances of $25$ Bohr and higher. This coincides 
with the value of $n=5.1$ observed here for the shorter separation distances. The excitation energy transfer rates extracted from Fig.\ \ref{fig:na4} are summarized in Table 
\ref{tab:na4}. 
\begin{table}[ht]
 \caption{Excitation energy transfer (EET) rate for the Na$_4$ cluster at the different separation distances.}
 \label{tab:na4}
 \centering
 \begin{tabular}{lcc}
  \hline
  R & $\kappa^{EET}$   \\
    (Bohr)& (fs)$^{-1}$    \\
  \hline
12.5 & 0.052  \\
15.0 & 0.022  \\
17.5 & 0.009  \\
  \hline
 \end{tabular}
\end{table}

\section{Summary and Conclusions}

We presented an extension of the subsystem DFT method to real-time TDDFT. In the new method, the electron density is evolved in time alongside a time dependent potential as a 
collection of time-dependent subsystem electron densities. The implementation is such that one can choose between evolving only one of the subsystems in time while keeping 
the other subsystems frozen at the initial time $t_0$ (uncoupled simulation) or evolving all subsystems simultaneously (coupled simulation). The former choice will only include 
the response of the excitations within the 
subsystem, while the latter will include all the couplings between all of the subsystems. 

The method was first applied to obtain optical absorption spectra of a benzene-fulvene 
dimer, where similar accuracy w.r.t. supramolecular DFT calculations were found as for the case of ground state FDE and LR-TDDFT FDE applications. For shorter distances, the 
optical absorption spectra start to diverge from the supramolecular calculations due to the known failure of NAKE functionals, while for intermediate and 
long distances the 
supramolecular results are reproduced with good accuracy. 

The choice between performing a coupled and an uncoupled calculation presented the opportunity to examine the effect of the excitation coupling between the subsystems on the 
reproduction of optical absorption spectra. A clear and consistent effect is seen on the interaction induced shift and the oscillator strength when omitting the full coupling of 
the subsystems, for the coupled excitations. Since we only examined a pair of nonidentical chromophores here, this effect is limited and we were able to confirm the previous 
findings\cite{neug2007} that subsystem LR-TDDFT is able to reproduce the supramolecular results for most systems while neglecting the intersubsystem coupling. 

As a second application, the real-time excitation energy transfer was studied in a Na$_4$ cluster, where only one of the Na$_2$ was subjected to a periodic electric field with a 
frequency of a coupled excitation energy. We found that the embedding potential is fully capable of transmitting the excitation energy between the subsystems and that the rate 
of the excitation energy transfer coincided with the F\"{o}rster theory prediction. 

The ability to represent the real-time dynamics of excitation energy transfer (EET) between subsystems is an important finding of this work. The ability to generate diabatic 
states for EET in a way that resembles most the true experiments (i.e., via an applied field) is of pivotal importance, complementing the linear-response theories. Future 
directions towards extending the method to period systems and to the inclusion of nuclear dynamics (Ehrenfest) are undergoing.



\end{document}